\documentclass[%
twocolumn,
 reprint,
superscriptaddress,
nofootinbib,
nobibnotes,
 amsmath,amssymb,
 aps,
]{revtex4-2}

\usepackage{graphicx,xcolor}
\usepackage{dcolumn}
\usepackage{bm}
\usepackage[breaklinks=true,colorlinks,citecolor=blue,linkcolor=blue,urlcolor=blue]{hyperref}
\usepackage[mathlines]{lineno}
\usepackage[caption=false, position=bottom]{subfig}
\usepackage{mathtools}
\usepackage{physics}
\usepackage[utf8]{inputenc}
\usepackage{mwe}
\usepackage{float}
\usepackage[font=small]{caption}
\begin{document}

\title{Quantum Chaos and  Circuit Parameter Optimization}

\author{Joonho Kim}
\affiliation{
 School of Natural Sciences, Institute for Advanced Study, Princeton, NJ 08540, USA.
}
 
\author{Yaron Oz}%
\affiliation{
Simons Center for Geometry and Physics, SUNY, Stony Brook, NY 11794, USA.
}
\affiliation{
 Raymond and Beverly Sackler School of Physics and Astronomy, Tel-Aviv University, Tel-Aviv 69978, Israel.
}

\author{Dario Rosa}%
\affiliation{
 Center for Theoretical Physics of Complex Systems, Institute for Basic Science (IBS), Daejeon - 34126, Korea
}

\begin{abstract}
We explore quantum chaos diagnostics of variational circuit states at random parameters and study their correlation with the circuit expressibility and the optimization of control parameters.
By measuring the operator spreading coefficient and the eigenvalue spectrum of the modular Hamiltonian of the reduced density matrix, we identify the universal structure of random matrix models in high-depth circuit states. We construct different layer unitaries corresponding to the GOE and GUE distributions and quantify their VQA performance. Our study also highlights a potential tension between the OTOC and BGS-type diagnostics of quantum chaos.

\end{abstract}

\date{\today}
\maketitle
\section{Introduction}


The random circuit model provides a framework of hybrid quantum/classical algorithms
for solving optimization and learning tasks, formulated as a search for the ground state of $k$-local Hamiltonians \cite{VQE2014, QAOA, Lucas_2014}.
Generic quantum gates create an entanglement between qubits. Entanglement is a valuable resource for achieving quantum advantage, but it also becomes a hurdle for successful optimization of circuit control parameters at the same time, specifically when random circuit states are much more highly entangled than the ground state of the Hamiltonian encoding the task \cite{McClean2018bp, cost-dep-bp,entanglement-bp,highdepth,Kim:2021ffs}.
Quantum information in such highly entangled states is scrambled, and
a successful adjustment of circuit parameters via local gradient search typically requires over-parametrization \cite{2021kkr, 2021qaoalandscape, Kim2, larocca2021theory}.

Quantum chaos \cite{BGS_seminal} is correlated with information scrambling \cite{larkin_ovchinnikov_OTOCs, Hayden_2007, Sekino_2008, shenker_stanford_2014, maldacena_shenker_stanford_2016}
and is in general a feature of interacting dynamical quantum systems \cite{dalessio_review_eth}.
One expects deep random circuit states to be generically chaotic. 
While the entanglement properties of a state can be quantified by entanglement entropies constructed from the eigenvalues
of its reduced density matrix, the quantum chaotic structure of the state is largely diagnosed 
by measures that depend on the level spacing of the reduced density matrix.
The aim of this work is to investigate the chaotic properties of random circuit states, with focus on the relationship between quantum chaos, circuit expressibility, and optimization performance.

A common diagnostic of quantum chaos that characterizes information scrambling is the operator spreading, which can be quantified by the 4-point out-of-time-order correlation function (OTOC) \cite{Sekino_2008, shenker_stanford_2014, kobryn_stanford_2021}.
One considers an operator at time
$t=0$, denoted by $\mathcal{O}(0)$, that acts on a small number of qubits and evolves it to:
\begin{equation}
\mathcal{O}(t) = U(t)^{\dagger} \mathcal{O}(0) U(t) \ ,
\label{evo}
\end{equation}
supported on a larger number of qubits at time step $t$.
This growth is typically ballistic with a characteristic velocity,
known as the butterfly velocity, reminiscent of the spread of classical chaotic trajectories. Considering a discrete time evolution driven by the random circuit unitary, this velocity depends on the circuit architecture, i.e., arrangement and type of quantum gate unitaries.
However, the operator growth is associated with  spectral values of the circuit reduced density matrix themselves, much like the entanglement entropy, but not their level spacings.
Let us denote the density matrix of the circuit state by $\rho_c$ and divide  
$n$ qubits of the quantum register into two subsets $A$ and $B$ of the equal size,  $n_A=n_B=\frac{n}{2}$. The modular Hamiltonian $H(\rho_A)$ of the reduced density matrix $\rho_A = \mathrm{Tr}_B \rho_c$ can be written as
\begin{equation}
\rho_A \equiv \frac{e^{-H(\rho_A)}}{Z_A} \ ,   
\label{mod}
\end{equation}
where $Z_A= \Tr_A e^{-H(\rho_A)}$ is the partition function of the modular Hamiltonian. It is indeed the eigenspectrum of $H(\rho_A)$ that encapsulates the entanglement and operator spreading properties of the quantum circuit.

There are by now accumulated pieces of evidence that chaotic properties
of Hamiltonian systems reveal themselves in the level spacing distribution of the Hamiltonian energy spectrum  \cite{haake_book}. 
This understanding can be extended to the eigenspectrum of the 
modular Hamiltonian, diagnosing the chaotic nature of a  \emph{quantum state} from its level spacing distribution \cite{Chen:2017yzn}.
We will explore various quantum chaos diagnostics as a function of the circuit depth
and show that deep circuit states exhibit the characteristics
of random matrix models.
Specifically, the level spacing distribution of the modular Hamiltonian, $r$-statistics,
and spectral form factor will manifest the universal structure of the Gaussian Orthogonal Ensemble (GOE) or Gaussian Unitary Ensemble (GUE), depending on types of quantum gates introduced in the random entangling circuit.

The paper is organized as follows.
In Section~\ref{sec:vqa}, we will describe the architecture of layered random circuits used for numerical simulation and briefly review the relationship between the
number of circuit layers and VQA performance. 
Section~\ref{sec:op_spread} will explore the connection between the operator spreading, a typical diagnostic of quantum chaos, and the optimization efficiency of control variables. We will then study in Section~\ref{sec:spectrum} the level spacing distributions of the modular Hamiltonians, $r$-statistics, and their spectral form factors at different circuit depths, showing that all diagnostics match those of random matrix ensembles in the high-depth regime. Section~\ref{sec:discussion} will conclude with discussion.

\section{VQA Performance}
\label{sec:vqa}
We begin with specifying the circuit architecture used in this paper and briefly review the relation between 
the entanglement made by random circuit unitaries and the optimization efficiency of variational quantum algorithms (VQAs) \cite{McClean2018bp, cost-dep-bp,entanglement-bp, highdepth,Kim:2021ffs,Kim2}.

\subsection{Circuit Architecture}

\begin{figure}[t]
\centering
\subfloat[\centering The circuit architecture]{
        \centering
        \label{fig:circuit_diagram_left}
        \includegraphics[height=1.85cm]{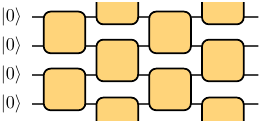}}
\hfill
\subfloat[\centering The quantum gates]{
        \centering
        \label{fig:circuit_diagram_right}
        \includegraphics[height=1.85cm]{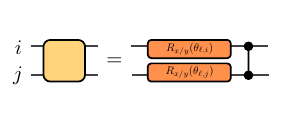}}    
\caption{The variational quantum circuit used in this paper. The $n$ qubits are arranged periodically, initialized in the state $|0\rangle^{\otimes n}$ and acted upon by a chain of the two-qubit unitary gates.}
\label{fig:circuit_diagram}
\end{figure}

Figure~\ref{fig:circuit_diagram} illustrates the variational circuit architecture assumed throughout our discussion. The $n$ quantum registers are arranged periodically, $i\simeq i+n$, and acted upon by a chain of two-qubit unitaries.
The unitaries are made of single-qubit gates $R(\theta)$ acting on all $n$ distinct qubits, 
followed by two-qubit entanglers:
\begin{align}
    CZ &= \text{diag}\left(+1, +1, +1, -1\right) \ \text{or}
    \label{eq:cz}\\
    CP &= \text{diag}\left(+1, +1, +1, +i\right) \ 
    \label{eq:cp}
\end{align}
that operate on all adjacent pairs of qubits. Every layer swaps the roles of odd/even qubits, alternating between controlling/controlled and controlled/controlling pairs.

In our numerical simulation, we will consider only two types of single-qubit gates. They are Pauli rotation gates along the $y$-axis
\begin{equation}
R_y(\theta) =  \exp(i  \sigma_y \theta) =    \begin{pmatrix}
\cos\theta & \sin\theta \\
-\sin\theta & \cos\theta
\end{pmatrix} \ ,
\label{eq:paulirot}
\end{equation}
which are real and orthogonal, and those along the $x$-axis
\begin{equation}
R_x(\theta) =  \exp(i  \sigma_x \theta )  =  \begin{pmatrix}
\cos\theta & i \sin\theta \\
-i \sin\theta & \cos\theta
\end{pmatrix} \ ,
\label{eq:paulirot2}
\end{equation}
which are complex-valued unitary matrices. All the rotation angles are randomly chosen from the uniform distribution $\mathcal{U}(0,2\pi)$ at circuit initialization. We will use the symbol $\theta_{\ell,i}$ to denote a specific angle that rotates the $i$'th qubit at the $\ell$'th layer, for $1\leq i \leq n$ and $1\leq \ell\leq L$.

We will consider four different types of layered circuits for numerical experiment. If the single-qubit rotations are all along the $x$ or $y$ axis, followed by the CZ entangler actions, the corresponding circuits will be called $R_x+ \text{CZ}$ or $R_y + \text{CZ}$. Two additional variants of circuit structures, dubbed as $R_x+ \text{CZ}+ R_y + \text{CZ}$ and $R_y + \text{CP}$, will be examined, where one-qubit gates at odd/even layers alternate between $R_x$/$R_y$ and where all CZ's are replaced with CP's, respectively. In general, different gate choices will lead to different entanglement and chaos properties.

\subsection{Optimization and Expressibility}

Random circuit states with a large number $L$ of layers  are typically highly entangled. Circuit expressibility, i.e., being able to represent generic states in the Hilbert space, can be achieved for sufficiently deep circuits. However, as quantum typicality makes the energy landscape of VQA Hamiltonians  flattened \cite{arrasmith2021equivalence, Kim2}, the circuit parameter optimization via local gradient search becomes more difficult with highly entangled circuits \cite{McClean2018bp, cost-dep-bp,entanglement-bp,highdepth,Kim:2021ffs}. A known remedy for the flattened energy landscape is overparametrization of variational ansatz \cite{kiani2020learning, 2021kkr, 2021qaoalandscape, larocca2021theory}, developing multiple steep directions that lead to the robust success of the gradient descent method \cite{Kim2}. This comes with a classical computational cost for storing and manipulating variables.

If a $k$-local Hamiltonian encodes the task to be solved, so that the corresponding ground state exhibits the area-law entanglement scaling, VQA will perform better with avoiding the region of quantum typicality \cite{McClean2018bp, cost-dep-bp}, including the saturation of bipartite entanglement entropy near the maximum value \cite{entanglement-bp,Kim:2021ffs}. 
A canonical example showing this relation is VQA with the one-dimensional Ising Hamiltonian made of the nearest-neighbor spin interaction coupled to a transverse magnetic field  \cite{sachdev2011}. Specifically, we will use two similar Hamiltonians that differ only in the direction of the external magnetic field:
\begin{align}
    H =  \sum_{i=1}^n \sigma_{z,i} \sigma_{z,i+1} + g\sum_{i=1}^n \sigma_{x,i}~~~\mathrm{or}~~~g\sum_{i=1}^n \sigma_{y,i}  \ ,
    \label{H}
\end{align}
and attempt to reach their ground states at $g=1$ by optimizing the circuit parameters at different $L$.

The search for optimal parameters that minimize the energy function will be conducted locally via the Adam optimization algorithm \cite{kingma2017adam}. It is a variant of the plain gradient descent that shows faster convergence in many circumstances, adjusting the step size at each iteration based on the moving average of gradients. 
We will choose its hyperparameters to be $(\alpha, \beta_1, \beta_2) = (0.05, 0.9, 0.999)$ in all numerical experiments. Per each run, we will allow enough time for convergence towards the ground state by waiting for $5000$ steps of the parameter update.

\begin{figure*}[t]
\centering
\subfloat[\centering $R_x$ + CZ]{
        \centering
        \includegraphics[height=4.0cm]{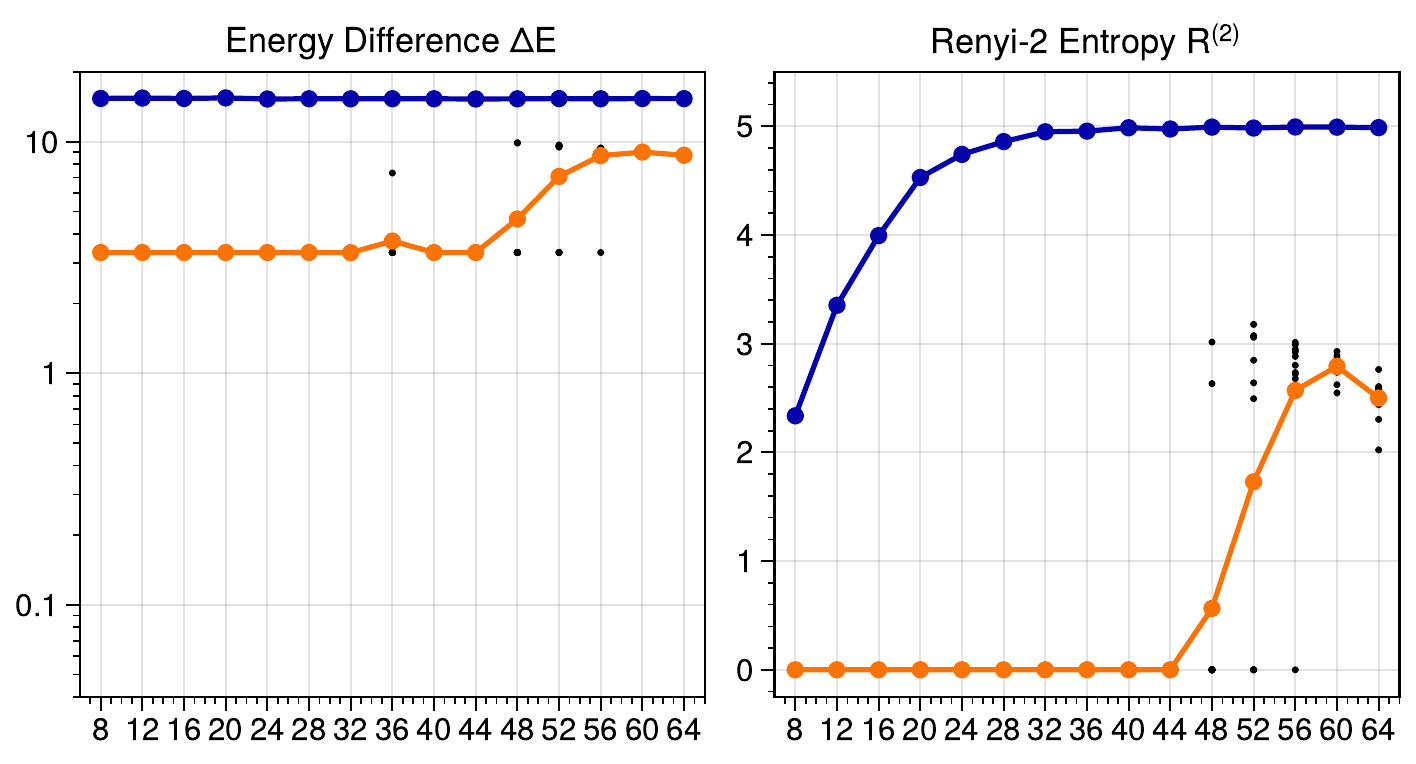}}
\subfloat[\centering $R_y$ + CZ]{
        \centering
        \includegraphics[height=4.0cm]{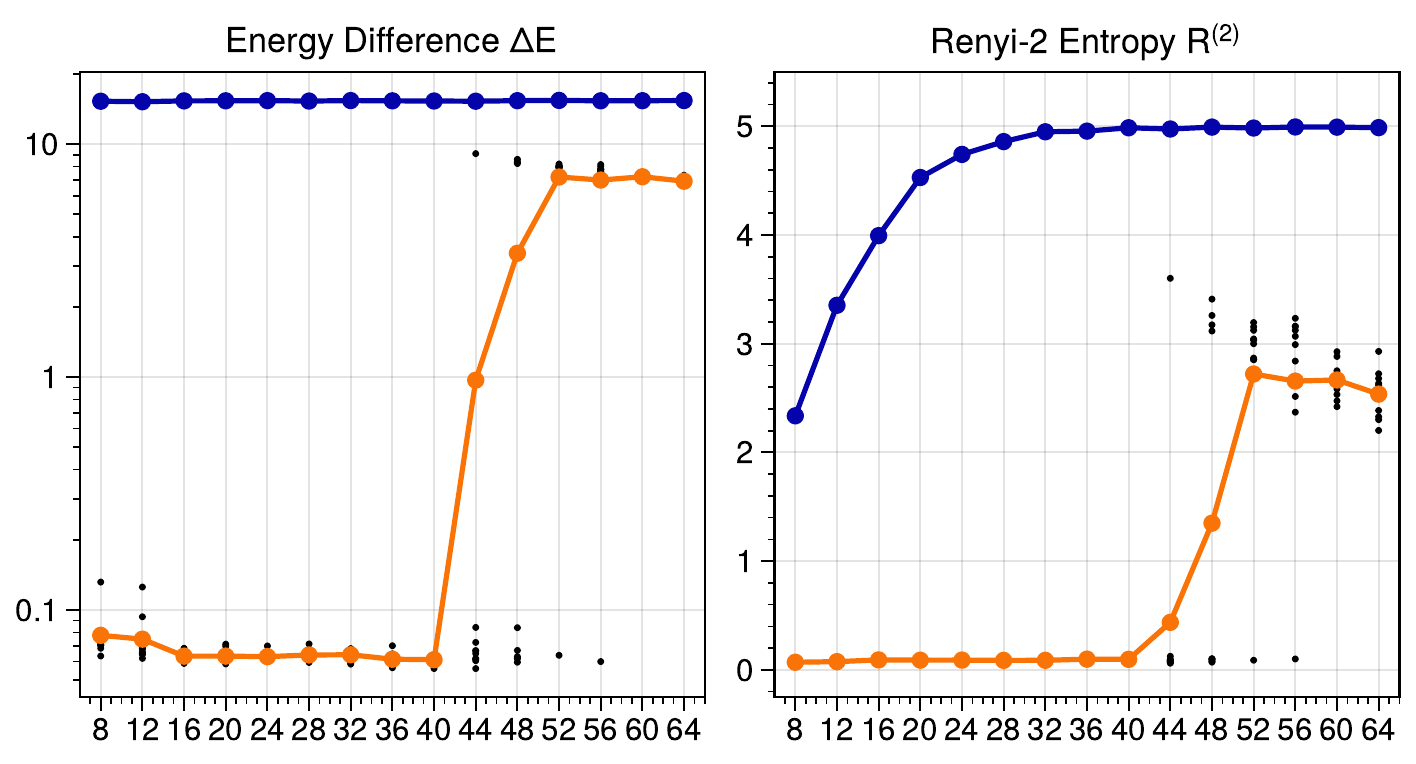}}  
        \hfill
\subfloat[\centering $R_x$ + CZ+$R_y$ + CZ]{
        \centering
        \includegraphics[height=4.0cm]{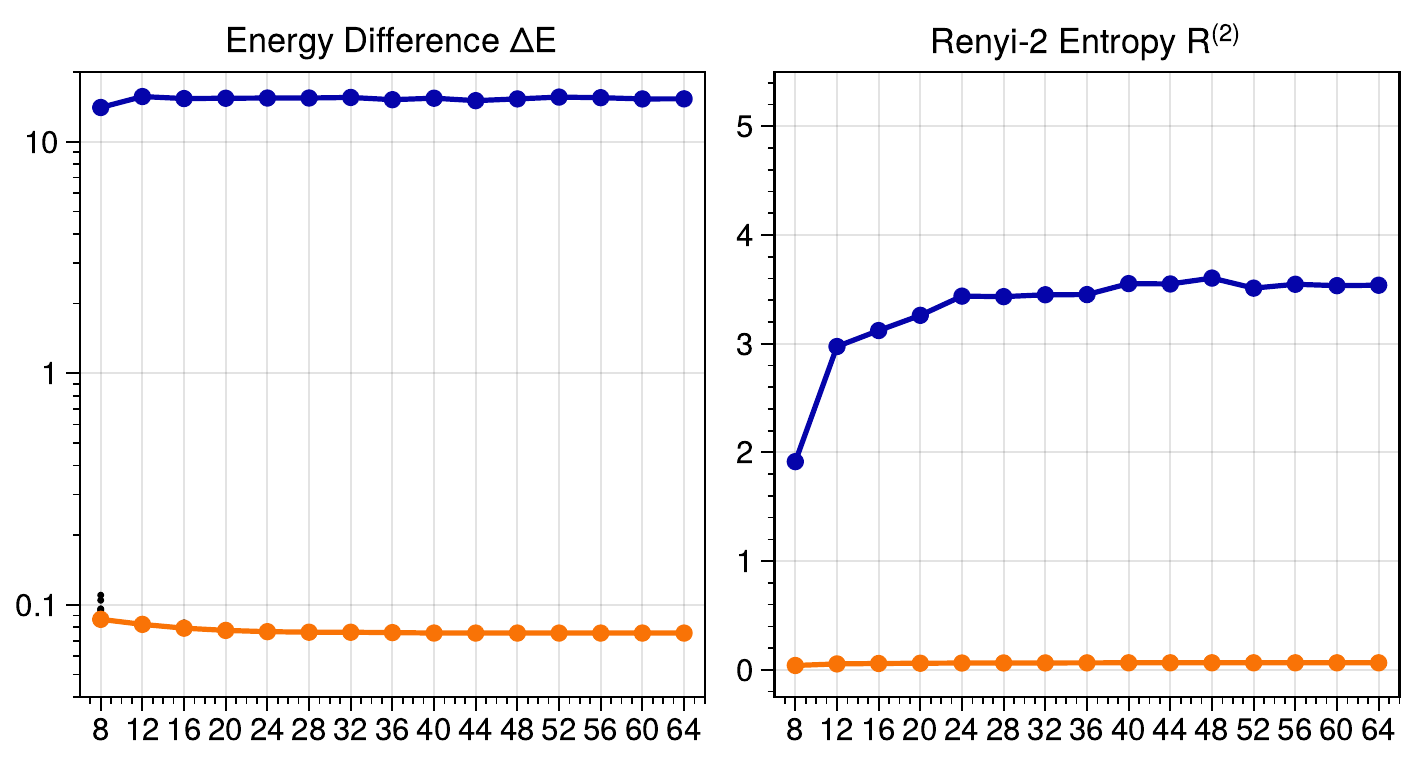}}  
\subfloat[\centering $R_y$ + CP]{
        \centering
        \includegraphics[height=4.0cm]{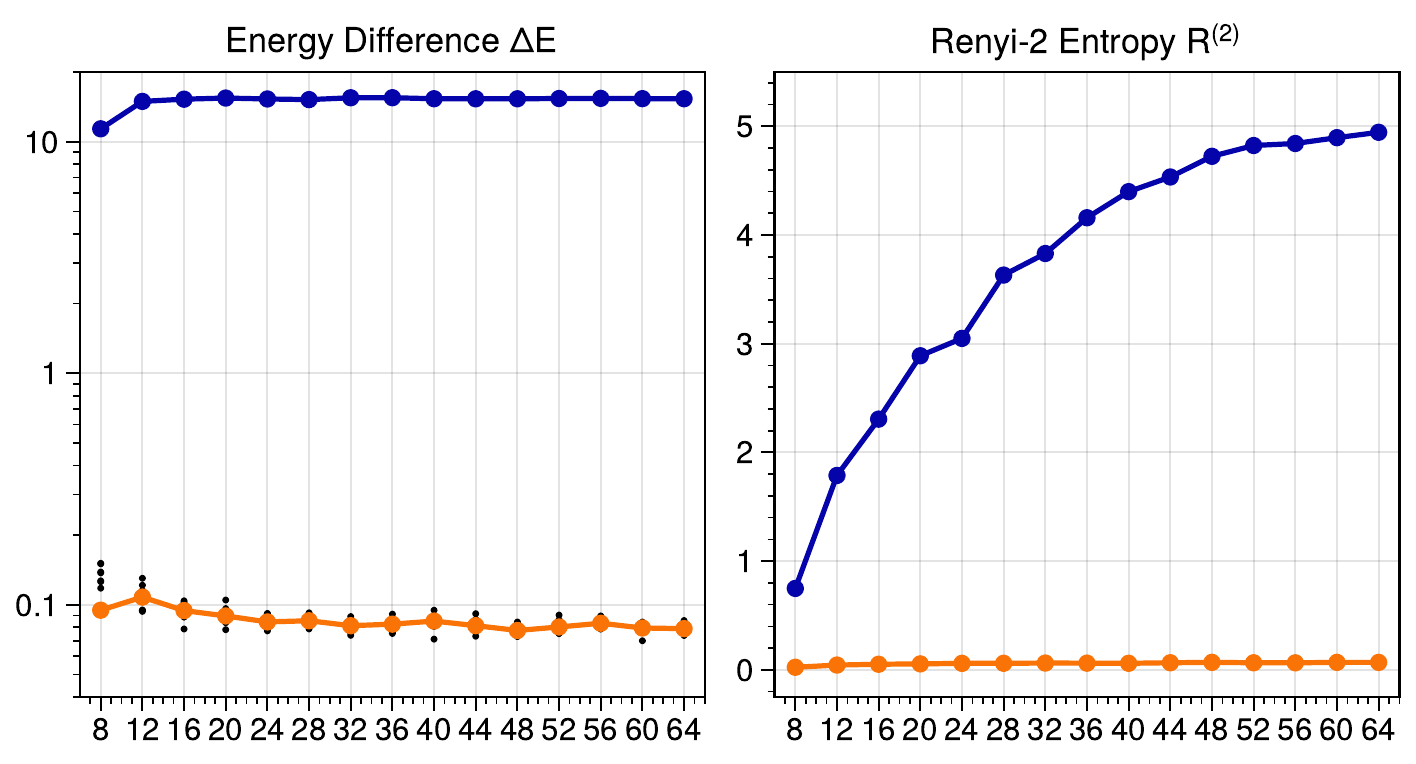}}  
\caption{Optimization curves for the Ising model coupled to an external field aligned to $x$-axis (\ref{H}) at $g=1$.
The plot on the left shows the energy difference from the ground state before (blue curves) and after (orange curves) optimization of the 
circuit parameters as a function of the number of circuit layers.
We see that the entanglement velocity (the slope of the curves) as well as the saturation 
plateau depend on the type of quantum gates.
The plot on the right shows the Renyi-2 entropy
before and after optimization of the 
circuit parameters as a function of the number of circuit layers.
(a) $R_x$ one-qubit rotation gate followed by entangling two-qubit $CZ$ gate.
(b) $R_y$ one-qubit gate followed by entangling two-qubit $CZ$ gate
(c) A sequence of $R_x$ one-qubit rotation gate, entangling two-qubit $CZ$ gate,
 $R_y$ and $CZ$.
 (d) $R_y$ one-qubit gate followed by entangling two-qubit $CP$ gate.} 
\label{fig:optim1}
\end{figure*}

\begin{figure*}[t]
\centering
\subfloat[\centering $R_x$ + CZ]{
        \centering
        \includegraphics[height=4.0cm]{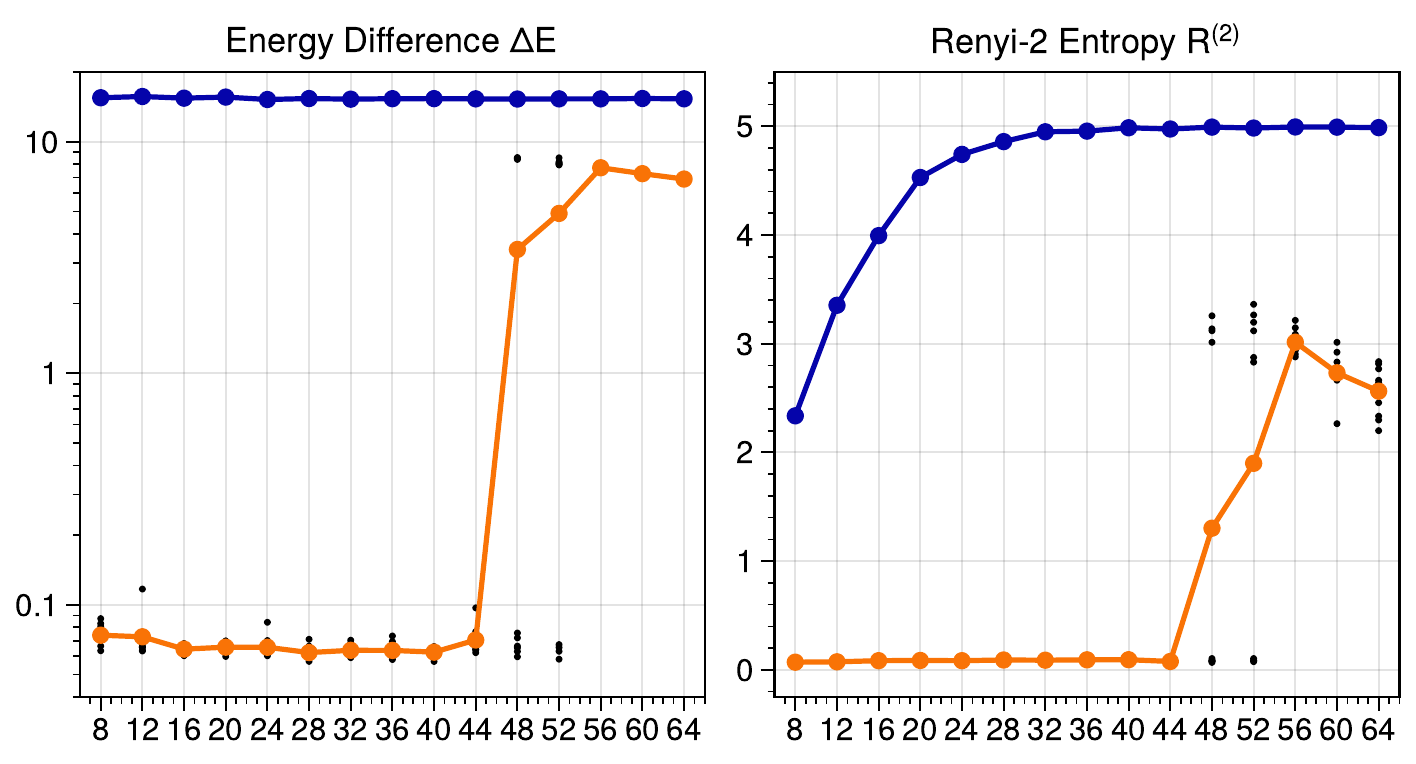}}
\subfloat[\centering $R_y$ + CZ]{
        \centering
        \includegraphics[height=4.0cm]{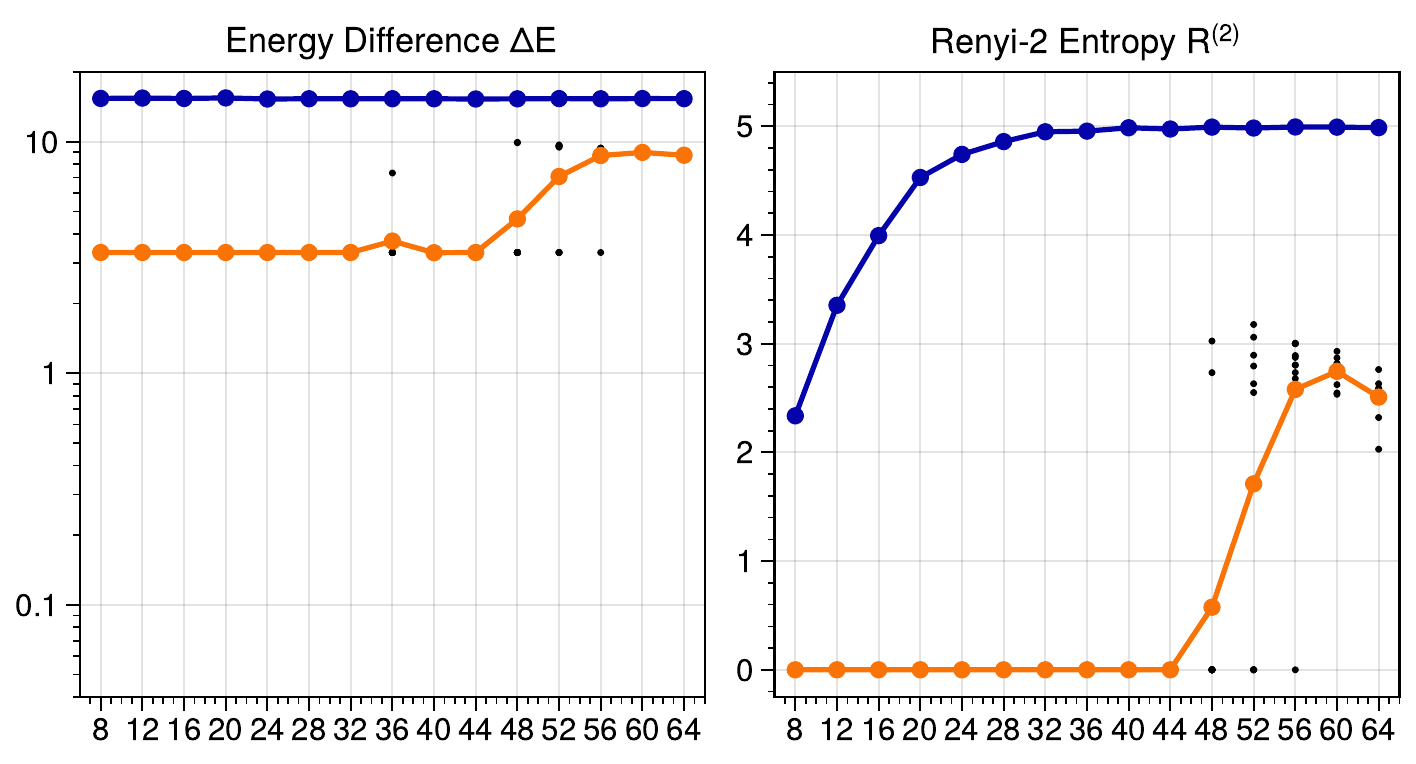}}  
        \hfill
\subfloat[\centering $R_x$ + CZ+$R_y$ + CZ]{
        \centering
        \includegraphics[height=4.0cm]{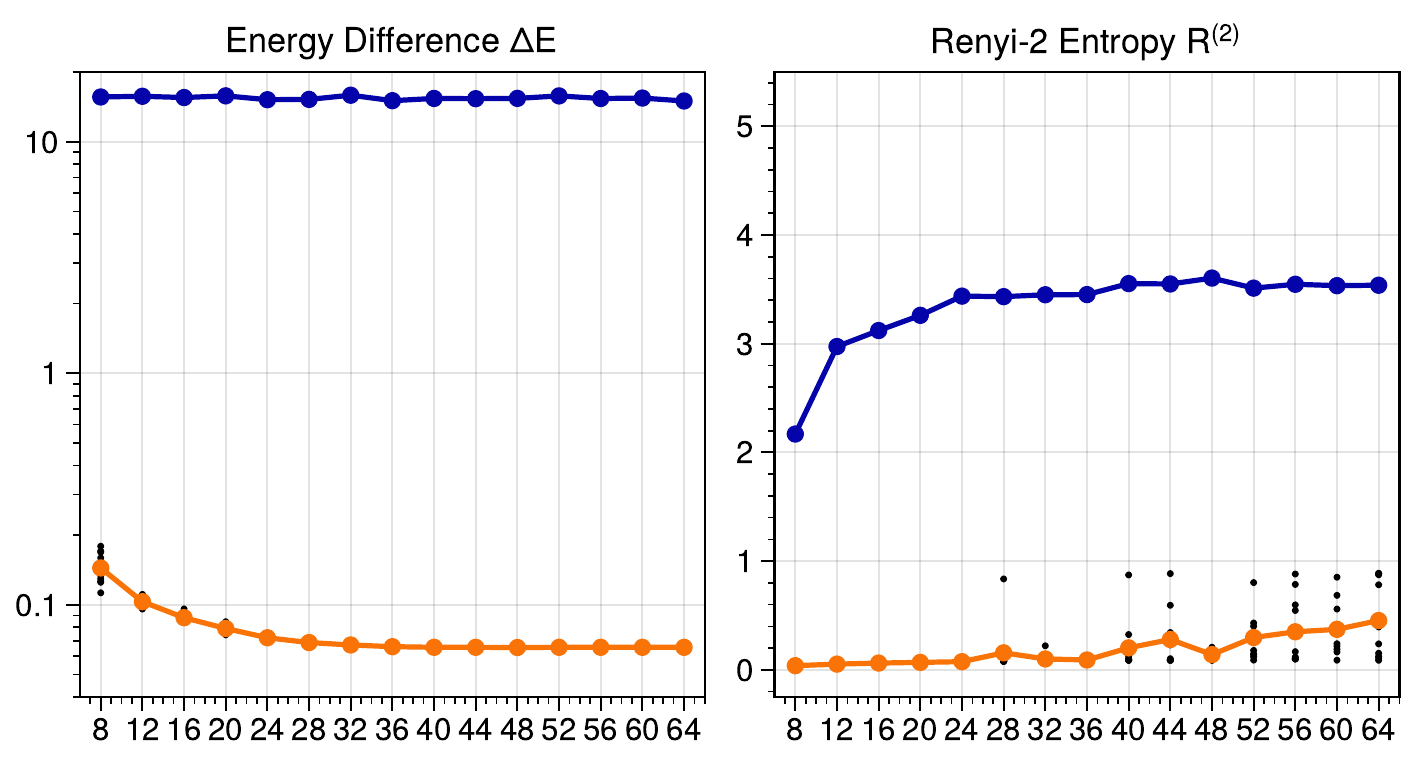}}  
\subfloat[\centering $R_y$ + CP]{
        \centering
        \includegraphics[height=4.0cm]{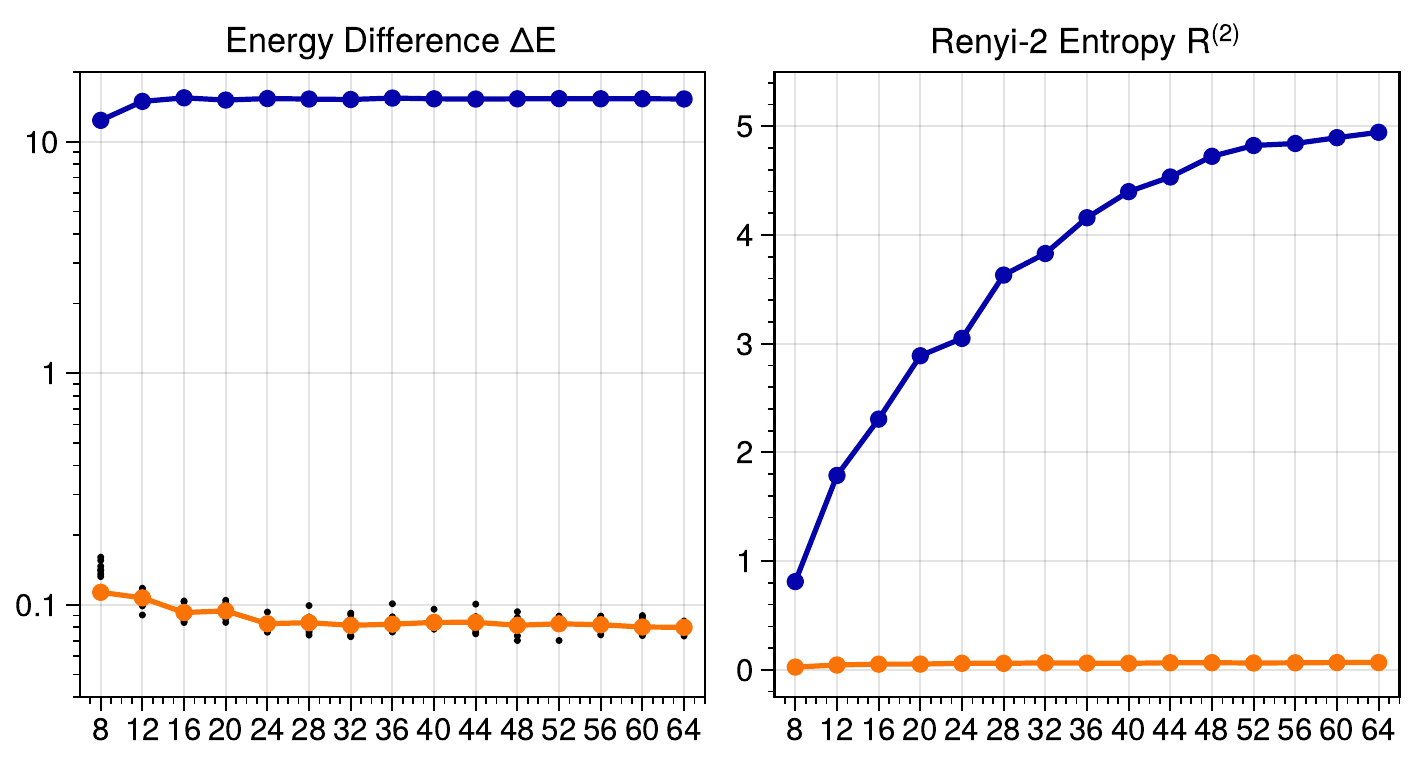}}  
\caption{Optimization curves for the Ising model coupled to an external field aligned to $y$-axis at $g=1$.
In comparison to Figure \ref{fig:optim1} we see that the circuit $R_y$ + CZ does not optimize well. It shows that 
the Renyi entropy can be insufficient to diagnose the optimization performance.
}
\label{fig:optim2}
\end{figure*}

From the collection of $10$ independent, repeated runs for each architecture and depth, the overall trend stands out: A high level of  expressibility, measured through the average Renyi entanglement entropy at random parameters saturating around its maximum possible value, has an adverse effect in reaching the ground state of the VQA Hamiltonian. See Figures~\ref{fig:optim1}~and~\ref{fig:optim2} for the outputs under four different choices of quantum gates. The orange/blue curves in their left panel display the energy gap from the ground state before/after the circuit parameter optimization as a function of $L$. Likewise, the orange/blue curves in the right panels represent the Renyi-2 entropy of the reduced density matrix 
\begin{align}
    {\cal R}^2_A = - \log  \tr_A \rho_A^2
\end{align}
obtained through partially tracing out $n/2$ qubits, before/after running the VQA optimization. 

It is notable that distinct gate choices lead to different entanglement growth and saturation values. For a particular circuit architecture dubbed as $R_x+ \text{CZ}+ R_y + \text{CZ}$, where one-qubit gate alternates between $R_x$ and $R_y$ at each layer, the entanglement curve converges to half the saturated value of other circuits. For $R_y + \text{CP}$ model that substitutes CZ entanglers with CP gates, the entanglement growth becomes considerably slower. They result in widening the depth window until approaching the maximum level of entanglement, in which the circuit parameter optimization can likely succeed. Also interestingly, we observe that $R_x+ \text{CZ}$/$R_y+ \text{CZ}$ models fail to reach the ground state of the Ising Hamiltonian coupled to the external field along  the $x$/$y$-axis, respectively.

\section{Operator Spreading}
\label{sec:op_spread}

Operator spreading serves as a diagnostic of the chaotic dynamics and information scrambling. It has been extensively studied in the context of random unitary  circuits, starting from \cite{Nahum_2018}. In this section, we will examine operator spreading as a function of the circuit depth $L$ that can be regarded as the time $t$ in discrete quantum systems.

Any Hermitian operator $\mathcal{O}(t)$ acting on $n$ qubit systems can be written in the Pauli string basis: 
\begin{align}
    \mathcal{O}(t) = \frac{1}{2^{n/2}}\sum_{j_1, \cdots, j_n} h_{j_1,\cdots,j_n}(t)\,  \sigma_{j_1}^{(1)}\otimes \cdots \otimes \sigma_{j_n}^{(n)} \ ,
\end{align}
where 
\begin{align}
    h_{j_1,\cdots,j_n}(t) \equiv\frac{1}{2^{n/2}} \text{Tr}(\sigma_{j_1}^{(1)}\otimes \cdots \otimes \sigma_{j_n}^{(n)}  \cdot \mathcal{O}(t)) \ .
\end{align}
Under the unitary time evolution (\ref{evo}),
\begin{eqnarray}
    &\text{Tr}&(\mathcal{O}(t)^\dagger \mathcal{O}(t))  = \text{Tr}(\mathcal{O}(0)^\dagger \mathcal{O}(0)) = \nonumber\\
  &=&  \frac{1}{2^n}\sum_{j_1, \cdots, j_n} \vert h_{j_1,\cdots,j_n}(t)\vert^2 = \text{constant} \ .
\end{eqnarray}
The size of the operator $\mathcal{O}(t)$ is defined as the size of the region where $\mathcal{O}(t)$ does not commute with
an operator $\sigma_a^{(x)}$ located at position $1 \leq  x \leq n$. It can be written as
\begin{eqnarray}
\mathcal{C}_a(x, t) &=& \frac{1}{2}\text{Tr} (\rho_\infty [\mathcal{O}(t), \sigma_a^{(x)}]^\dagger [\mathcal{O}(t), \sigma_a^{(x)}])  = \nonumber\\
&=& \frac{1}{2}\text{Tr} ( [\mathcal{O}(t), \sigma_a^{(x)}]^\dagger [\mathcal{O}(t), \sigma_a^{(x)}]) \nonumber \\
&=& \sum_{\substack{j_1,\cdots,j_n \\j_x \neq 0, a}} 2 \vert h_{j_1,\cdots,j_n}(t)\vert^2 \ .
\label{c}
\end{eqnarray}
We numerically measure it with $a=y$, where the operator ${\cal O}(0)$ is the Pauli-$x$ matrix located at $x=n/2$.

\begin{figure*}
\subfloat[\centering $R_x$ + CZ]{
\centering
        \includegraphics[height=4.0cm]{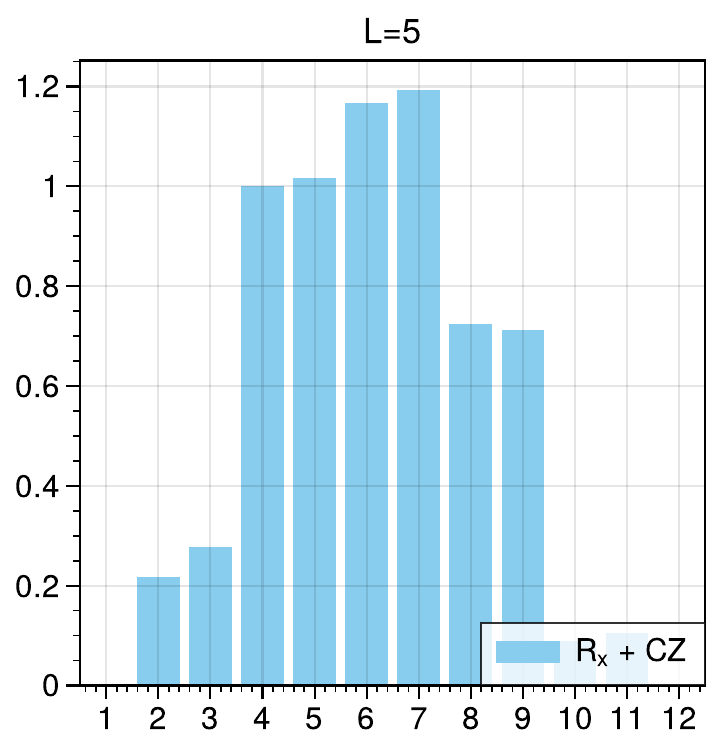}
        \includegraphics[height=4.0cm]{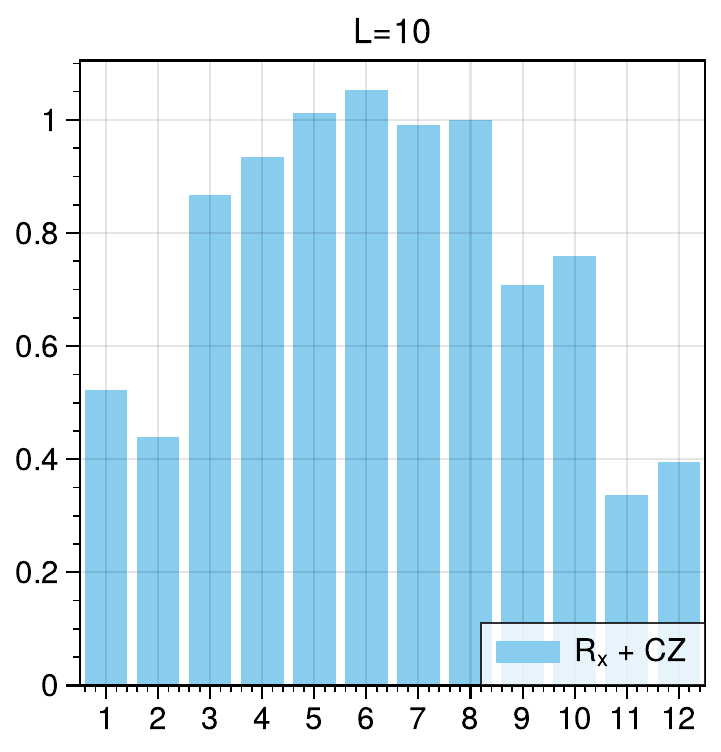}
        \includegraphics[height=4.0cm]{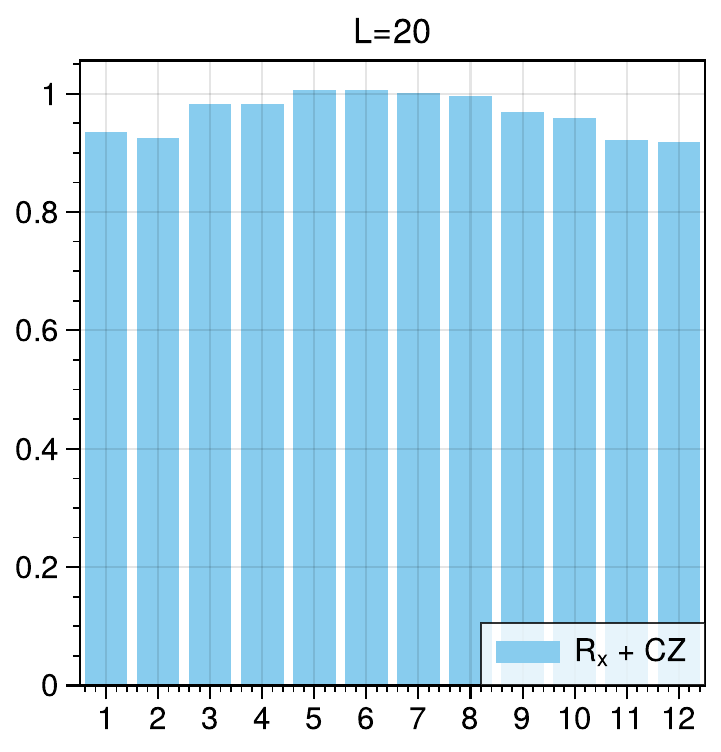}
        \includegraphics[height=4.0cm]{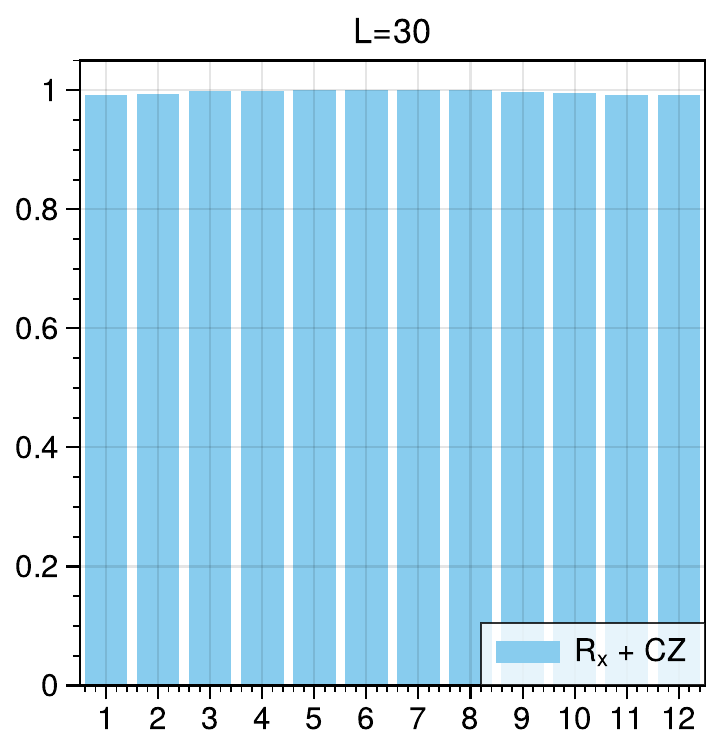} }
        
\subfloat[\centering $R_x$ + CZ+$R_y$ + CZ]{
        \includegraphics[height=4.0cm]{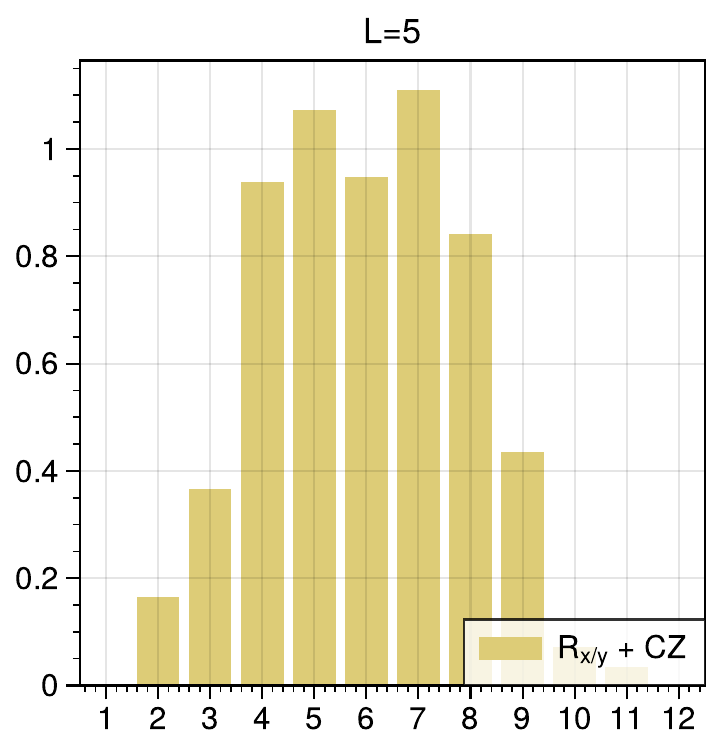}
        \includegraphics[height=4.0cm]{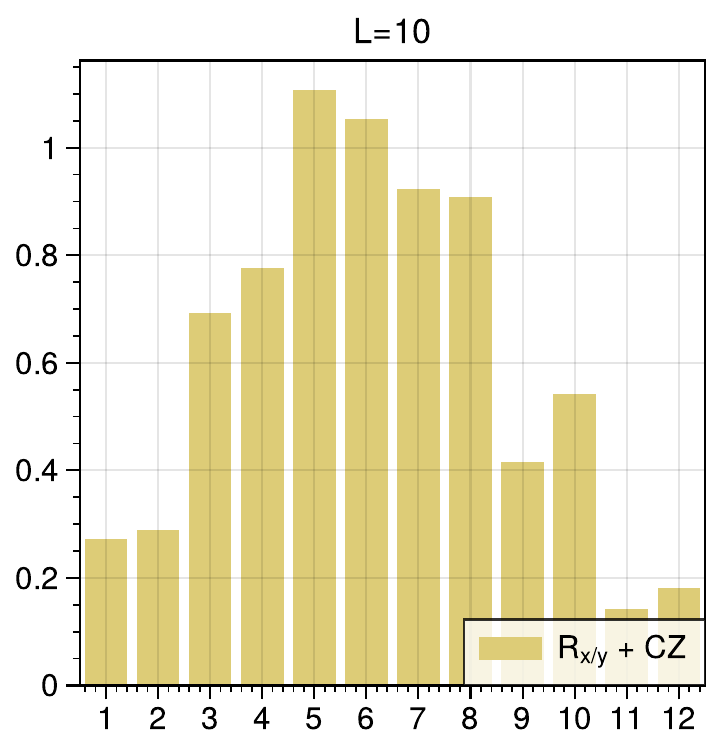}
        \includegraphics[height=4.0cm]{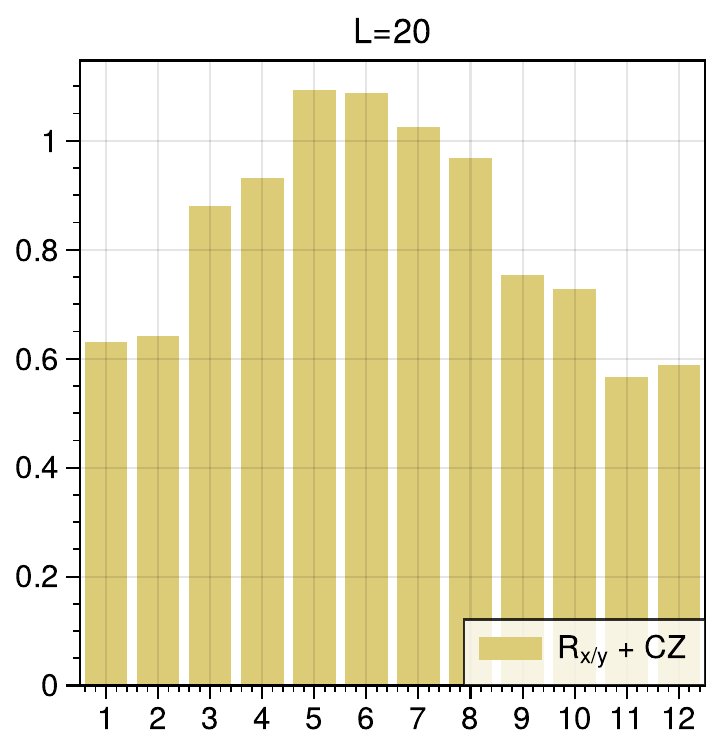}
        \includegraphics[height=4.0cm]{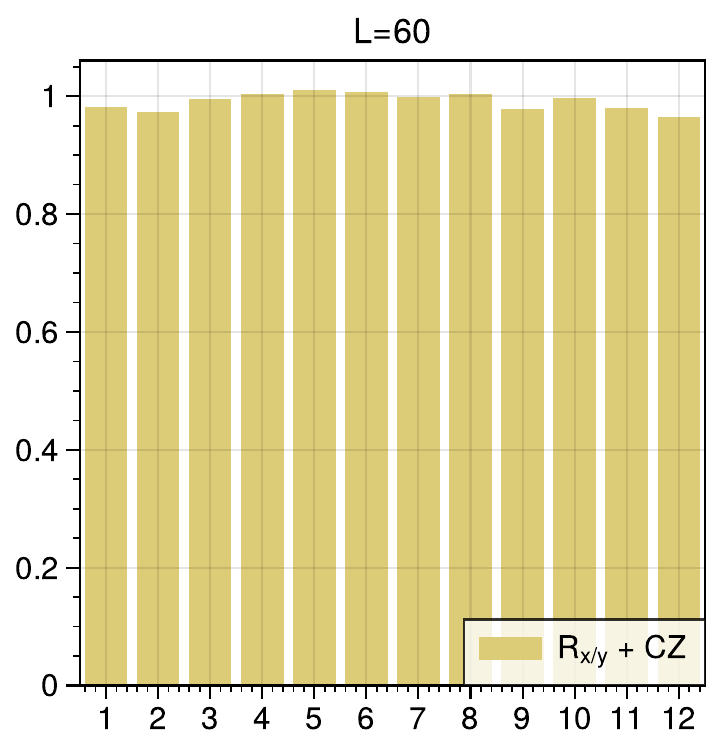}
        }
        
\subfloat[\centering $R_y$ + CZ]{
\centering
        \includegraphics[height=4.0cm]{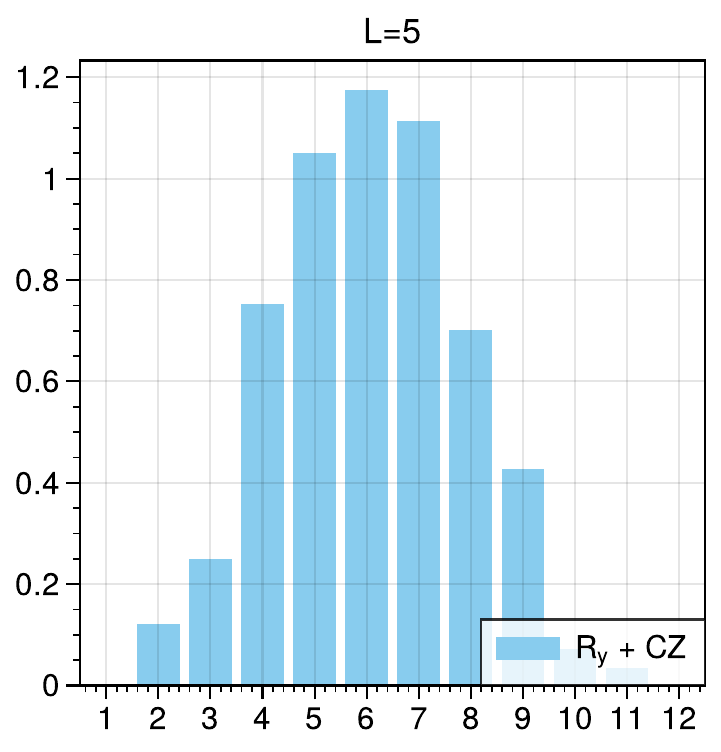}
        \includegraphics[height=4.0cm]{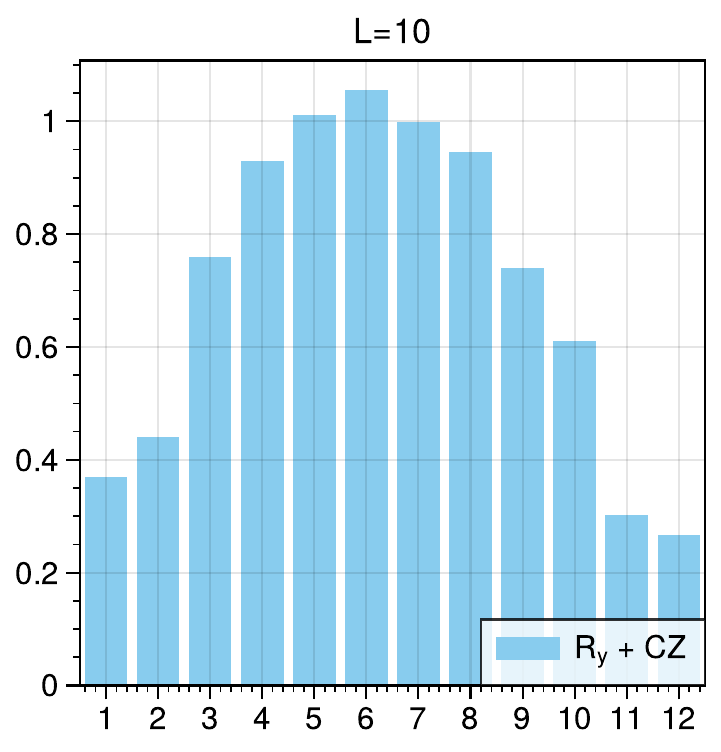}
        \includegraphics[height=4.0cm]{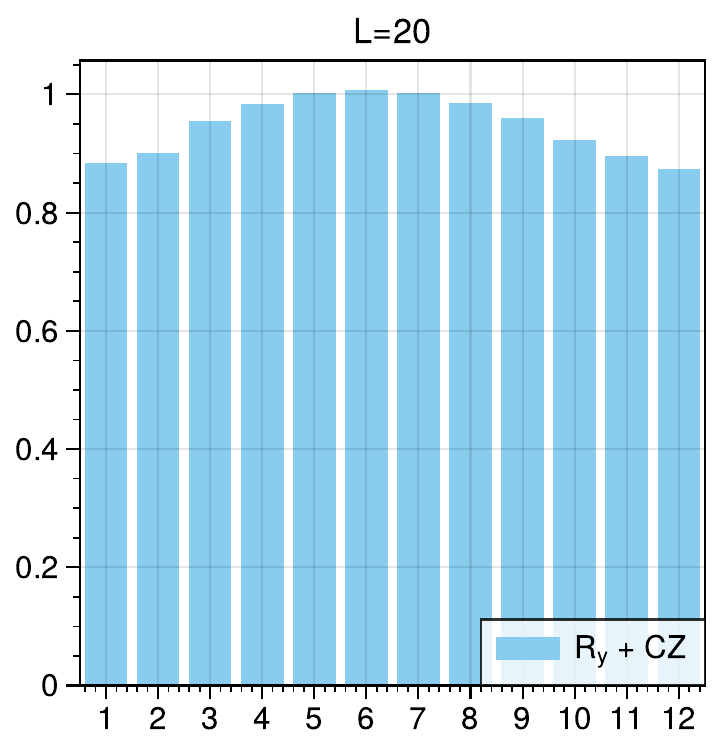}
        \includegraphics[height=4.0cm]{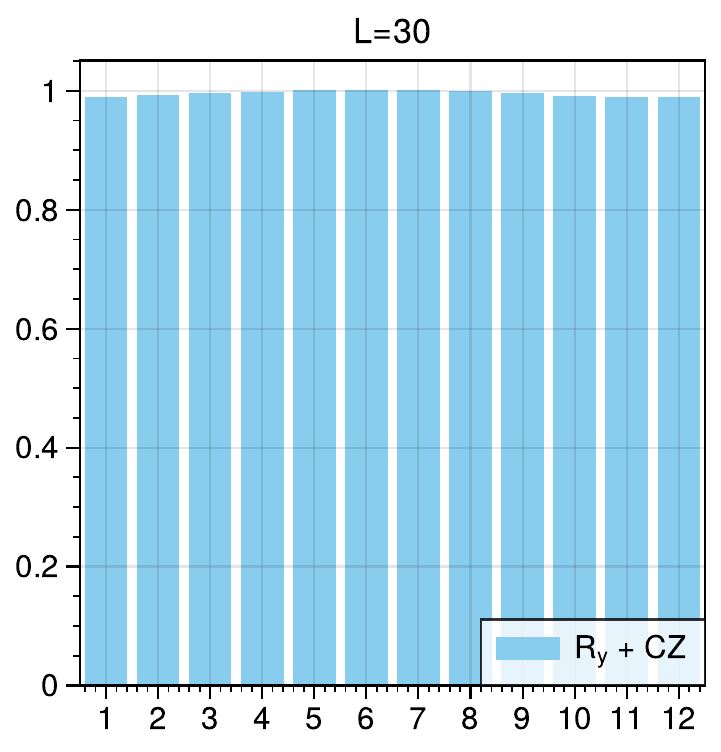}}

\subfloat[\centering $R_y$ + CP]{
        \includegraphics[height=4.0cm]{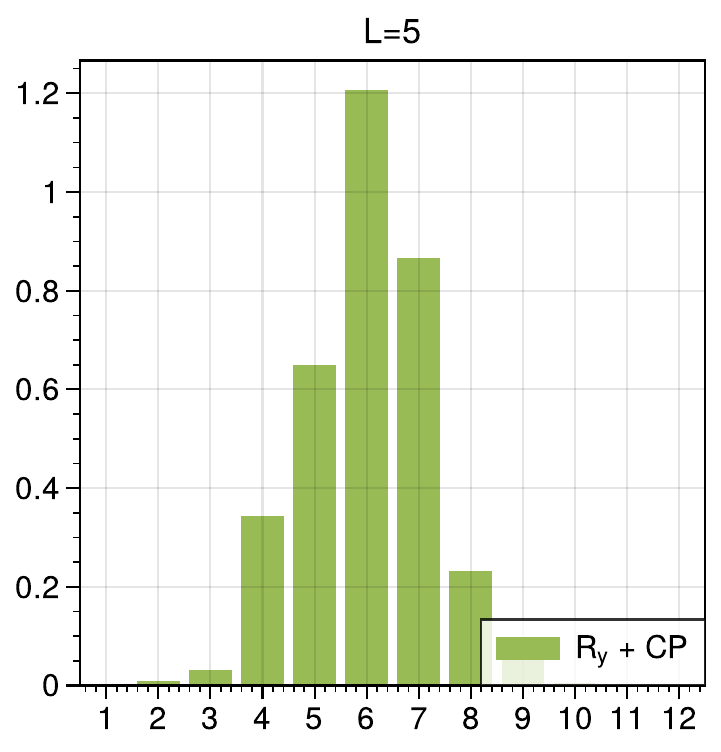}
        \includegraphics[height=4.0cm]{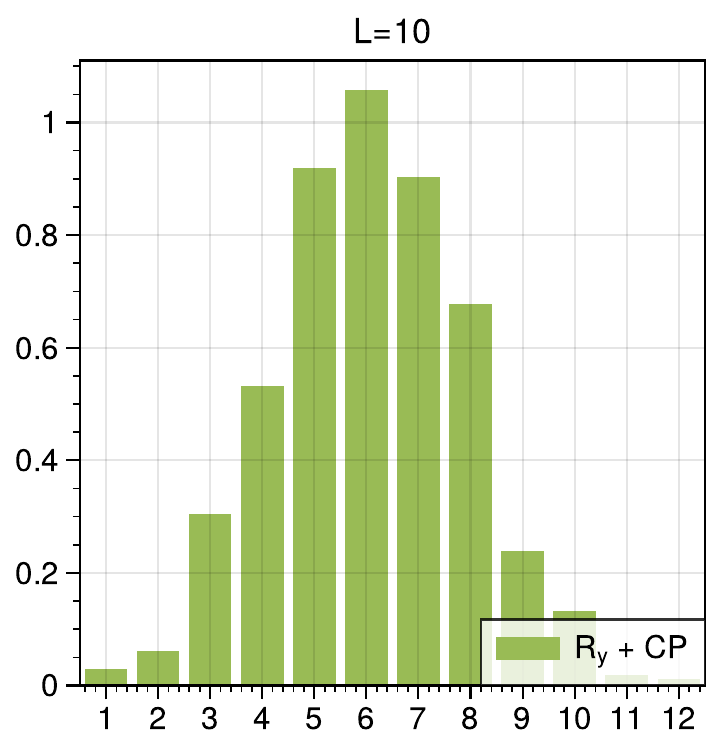}
        \includegraphics[height=4.0cm]{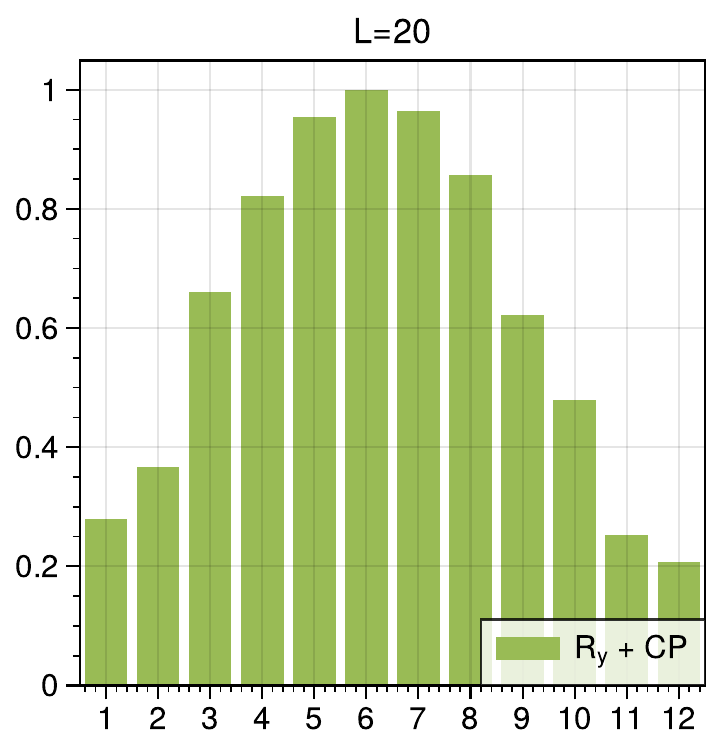}
        \includegraphics[height=4.0cm]{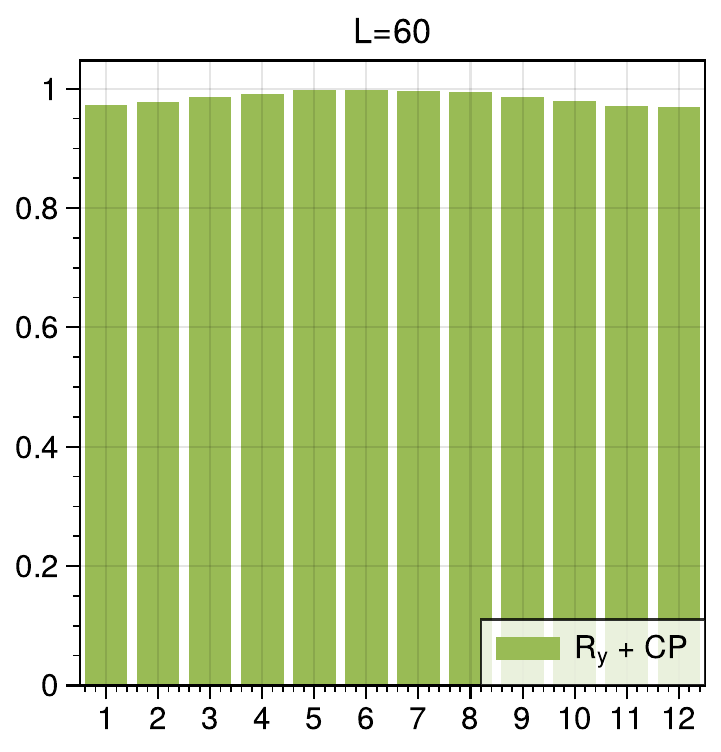} }
\caption{The operator spreading $\mathcal{C}(x,t)$ for a system of $n=12$ qubits 
as a function of $1 \leq x \leq n=12$ for different 
number of layers $L$. The different circuit architectures are:
(a) $R_x$ one-qubit rotation gate followed by entangling two-qubit $CZ$ gate.
(b) $R_y$ one-qubit gate followed by entangling two-qubit $CZ$ gate
(c) A sequence of $R_x$ one-qubit rotation gate, entangling two-qubit $CZ$ gate,
 $R_y$ and $CZ$.
 (d) $R_y$ one-qubit gate followed by entangling two-qubit $CP$ gate.
 We see a clear correlation between the operator spreading and the entanglement
 measures of the circuit in Figures \ref{fig:optim1} and \ref{fig:optim2}.}
 \label{osmeasure}
\end{figure*}

\begin{figure}[H]
\centering
\includegraphics[height=5cm]{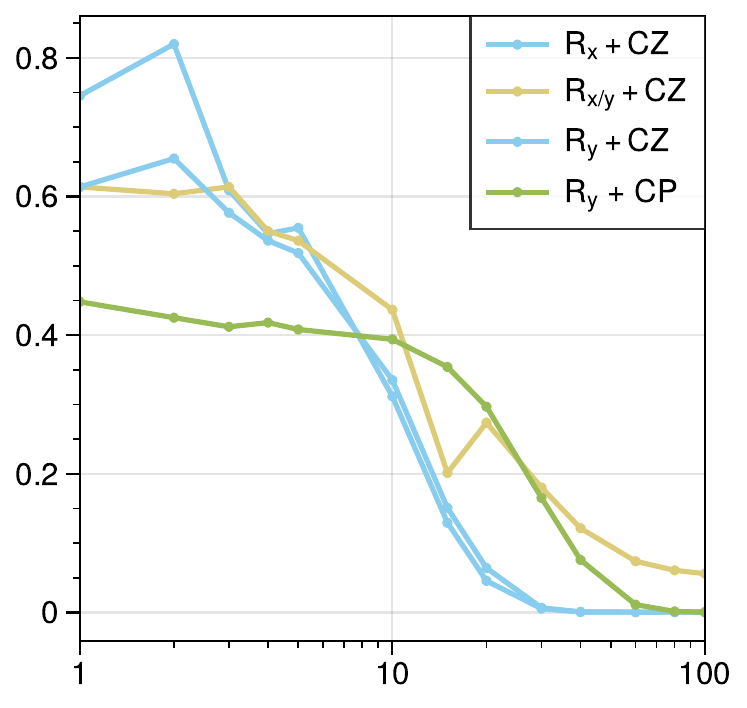}
\caption{Standard deviation of $C(x,t)$ over all lattice sites $x$ as a function of the circuit depth $t$.
We see that even at a large number of layers, there is no complete spreading for the $R_x$ + CZ +$R_y$ + CZ circuit. This is consistent with the saturation behaviour of the circuit entanglement in Figure \ref{fig:optim1} (c) and in Figure \ref{fig:optim2} (c).}
\label{SD}
\end{figure}

Figure~\ref{osmeasure} visualizes the operator spreading coefficient $\mathcal{C}_y(x, t)$ at different times $t=L$ and positions $1 \leq x \leq n$, averaged over 50 random circuit instances in a system of $n=12$ qubits. For comparison of the operator spreading pattern across different quantum circuit architectures, we also draw in Figure \ref{SD} the standard deviation of $\mathcal{C}_y(x, t)$ over all $1 \leq x \leq n$ as a function of $L$. We observe that the $R_x + \text{CZ}$ and $R_y + \text{CZ}$ unitaries reach saturation around $L\gtrsim 30$, while $R_y + \text{CP}$ takes $L\gtrsim 60$ for complete spread. Furthermore, there is no complete spreading under the $R_x + \text{CZ} +R_y + \text{CZ}$  unitaries even with a large number of circuit layers $L$. These behaviors are all consistent with the entanglement growth pattern of random circuit states illustrated in Figures~\ref{fig:optim1}~and~\ref{fig:optim2}, showing a clear correlation between two distinct quantities.

\section{Spectral diagnostics of quantum chaos}
\label{sec:spectrum}

The Bohigas-Giannoni-Schmit (BGS) conjecture \cite{BGS_seminal, Guhr_1998} associates 
the quantum chaotic properties of a system with the correlations between its energy levels.
Chaotic Hamiltonians exhibit level correlations in agreement with the predictions of random matrix theory (RMT) \cite{mehta_book}.
Adjacent eigenvalues show level repulsion and, at larger energy scales, signals of spectral rigidity.

In this section, we will apply three quantum chaos diagnostics, i.e., the level spacing distribution, the $r$-statistics and the spectral form factor, to the modular Hamiltonian (\ref{mod}) of quantum circuits at varying depth $L$.
The first two diagnoses focus on small energy scales and therefore can determine the presence of level repulsion, one of the most robust indications of quantum chaos \cite{haake_book}. On the other hand, the spectral form factor probes larger energy scales and is mostly a quantifier for spectral rigidity. 

A particular care needs to be taken in analysing the eigenvalues of the reduced density matrices $\rho_A$, since they show unavoidable numerical errors.
In order to control the effect of the numerical errors, we have adopted a robust phenomenological procedure, which makes use of the fact that all the eigenvalues of $\rho_A$ must be non-vanishing by definition.
Let us denote by $\lambda_{\min}$ the minimum negative eigenvalue among $N$ ensemble realizations for a given value of $L$.
To make sure that we consider only eigenvalues of $\rho_A(L)$ that are not affected by the numerical precision, we impose a cutoff on the spectra by considering only the eigenvalues satisfying the bound:
\begin{equation}
    \lambda_i \geq 10  |\lambda_{\min} | \ .
\end{equation}
Such cutoff, when applied at small values of $L$, removes most of the eigenvalues of $\rho_A(L)$
as  most of the eigenvalues are zero at small $L$. However, this is not the case for larger $L$, when the RMT structure is clearly visible. 
The procedure ensures that the eigenvalues kept are robust and not significantly affected by the numerical precision.
Out of the significant eigenvalues of $\rho_A(L)$, we compute energy levels $E_i$ of the modular Hamiltonian $H(\rho_A)$.

In the next subsections, we will consider only meaningful energy levels, $E_i$, obtained from the above procedure.

\subsection{Level Spacing Distribution}
\label{sec:level_spacings}

Roughly speaking, the level spacing distribution measures the probability density for two adjacent eigenvalues to be in the energy distance $s$, in units of the mean level spacing $\Delta$.
The procedure for normalizing all distances in terms of the local mean level spacing is often referred to as unfolding.
We unfold the spectrum of the modular Hamiltonian $H(\rho_A)$ by using the following algorithm:
\begin{enumerate}
    \item Arrange non-degenerate energy levels, $E_i$, of a modular Hamiltonian $H(\rho_a)$ in ascending order.
    \item Compute the staircase function $S(E)$ that enumerates all eigenstates of  $H(\rho_a)$ whose eigenvalues 
    are smaller than or equal to $E$.
    \item Fit a smooth curve that we denote by $\Tilde{\rho}(E)$ to the staircase function. To be specific, we used a $12$-th order polynomial as the smooth approximation.
    \item Rescale the energy levels $E_i$ as follows:
    \begin{equation}
        \label{eq:unfolded_spectrum}
        E_i \to e_i = \Tilde{\rho}(E_i) \, .
    \end{equation}
    \item By construction, the unfolded energy levels $e_i$ must show an approximately uniform distribution with mean level spacing 1. This can be used to ensure if the above procedure has been successful, i.e., by plotting the unfolded levels and checking  the flatness of the distribution.
\end{enumerate}

Having obtained the unfolded spectrum, we compute their  level spacing, $s_{i} = e_{i+1}-e_i$, and draw the probability density function $p(s)$ for having two neighbouring eigenvalues separated at a distance $s$. The level spacing distribution serves as a diagnostic for quantum chaos in Hamiltonian systems. It captures information about the short-range spectral correlations. 
It thus demonstrates the presence of level repulsion, {\it i.e.} whether $p(s)\rightarrow 0$ as $s\rightarrow 0$, which is a common characteristic of random matrix ensembles and particularlly chaotic Hamiltonians.

The level spacing distribution $p(s)$ for integrable systems follows the Poisson distribution 
\begin{equation}
 p(s) = e^{-s} \ ,   
 \label{po}
\end{equation}
while for chaotic systems it takes the following form
\begin{equation}
  p_{\beta}(s) = \frac{s^{\beta} e^{-b_{\beta}s^2}}{\Gamma(\frac{1+\beta}{2}) }  \ ,
  \label{Wigner}
\end{equation}
where $\beta$ depends on which universality class of random matrices the chaotic Hamiltonian belongs to \cite{mehta_book}: $\beta=1$ for the Gaussian Orthogonal Ensemble (GOE), $\beta=2$ for the Gaussian Unitary Ensemble (GUE), and $\beta = 4$ for the Gaussian Symplectic Ensemble (GSE).

\begin{figure*}[t]
\centering
\subfloat[\centering $R_x$ + CZ]{
        \centering
        \includegraphics[height=4.0cm]{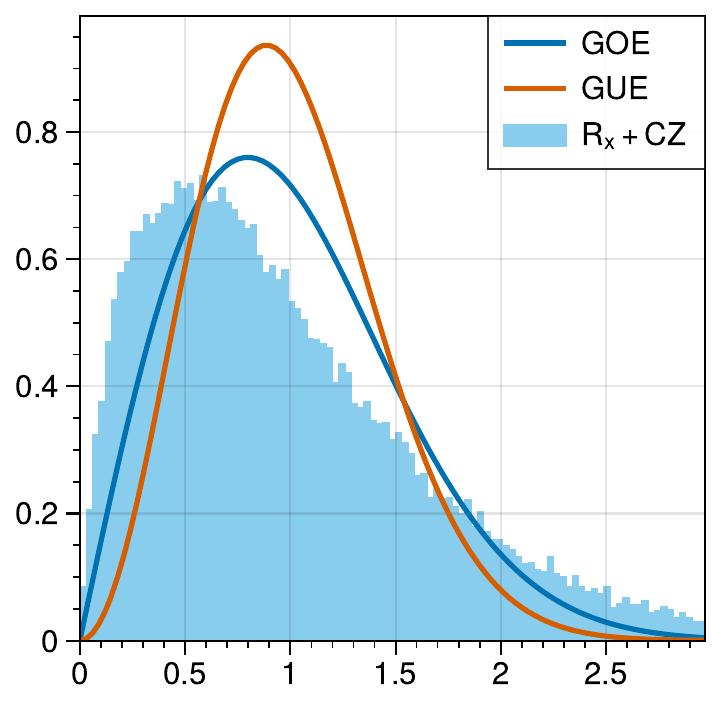}}
        \hfill
\subfloat[\centering $R_y$ + CZ]{
        \centering
        \includegraphics[height=4.0cm]{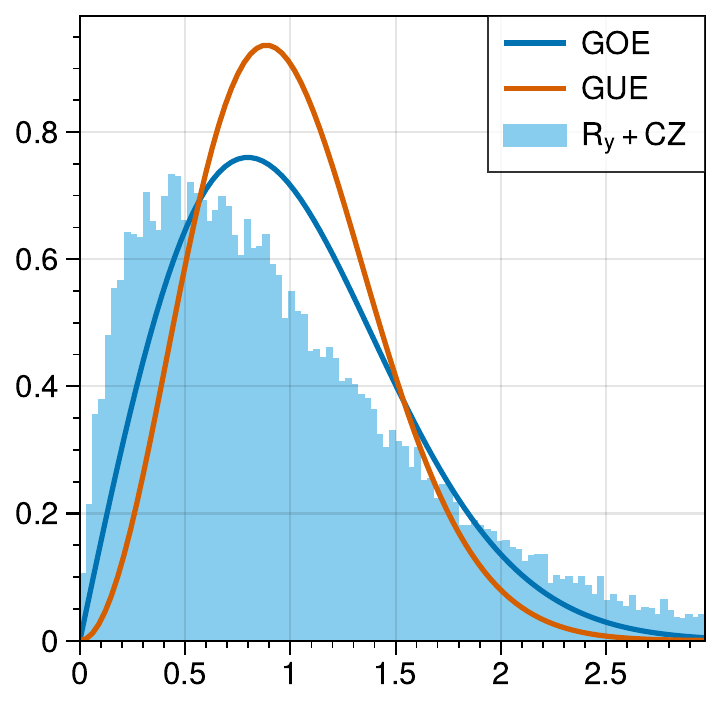}}  
        \hfill
\subfloat[\centering $R_x$ + CZ+$R_y$ + CZ]{
        \centering
        \includegraphics[height=4.0cm]{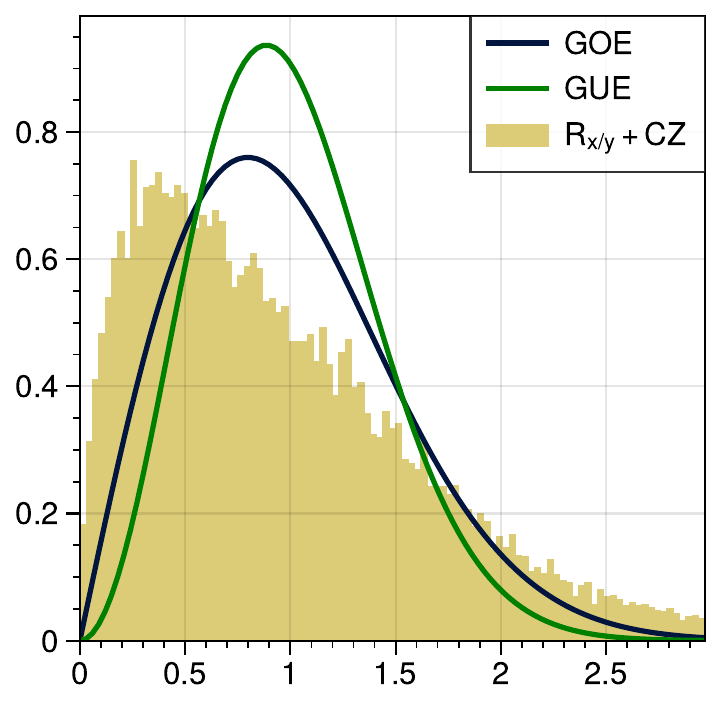}}  
        \hfill
\subfloat[\centering $R_y$ + CP]{
        \centering
        \includegraphics[height=4.0cm]{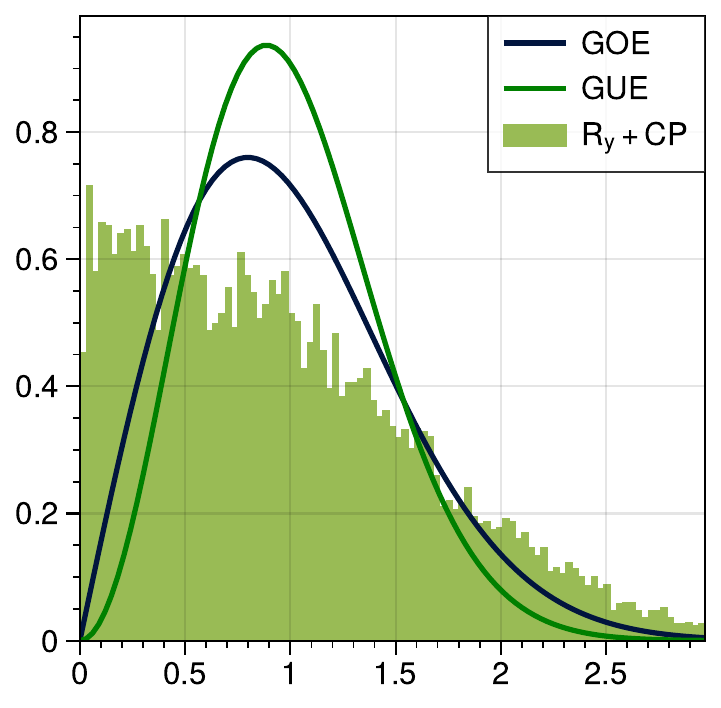}}
\caption{Level spacing distributions at $L=10$ number of layers. The distributions differ from those of 
RMTs.}
\label{fig:ls10}
\end{figure*}

\begin{figure*}[t]
\centering
\subfloat[\centering $R_x$ + CZ]{
        \centering
        \includegraphics[height=4.0cm]{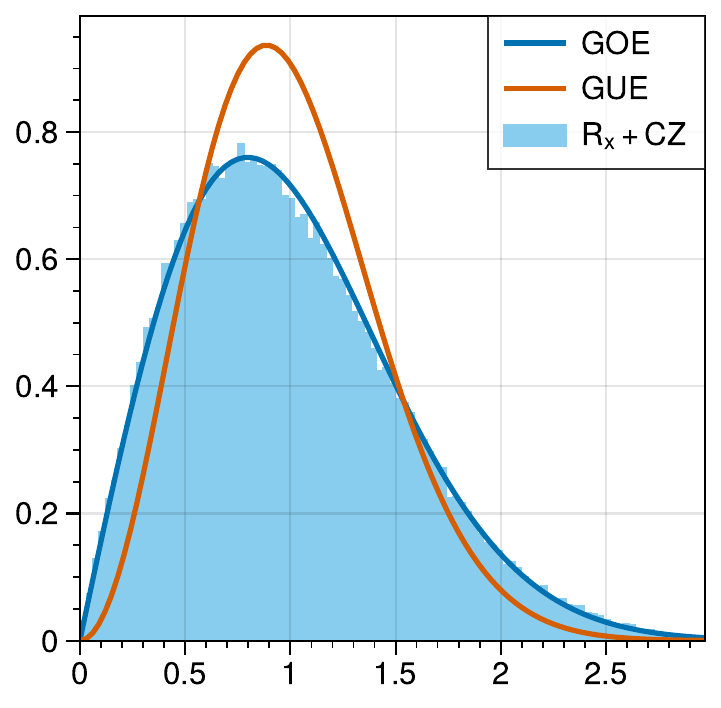}}
        \hfill
\subfloat[\centering $R_y$ + CZ]{
        \centering
        \includegraphics[height=4.0cm]{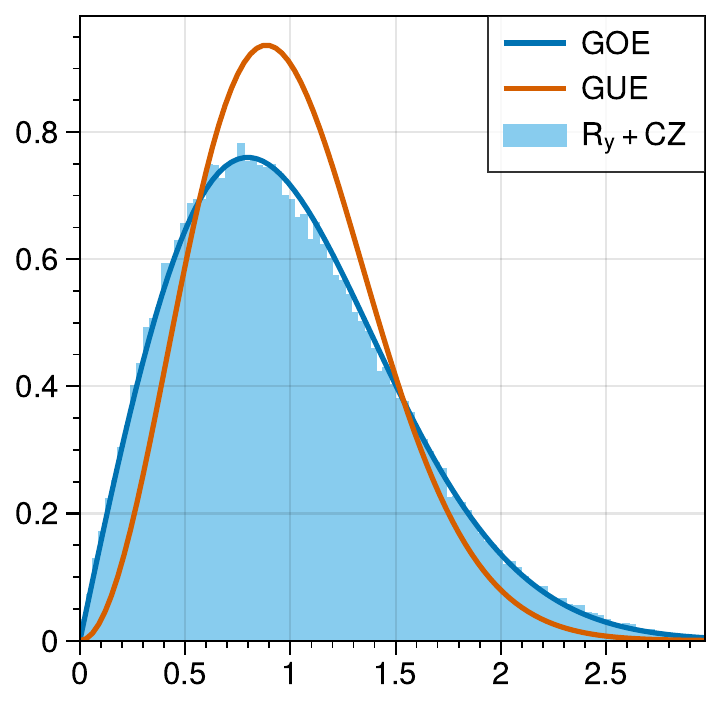}}  
        \hfill
\subfloat[\centering $R_x$ + CZ+$R_y$ + CZ]{
        \centering
        \includegraphics[height=4.0cm]{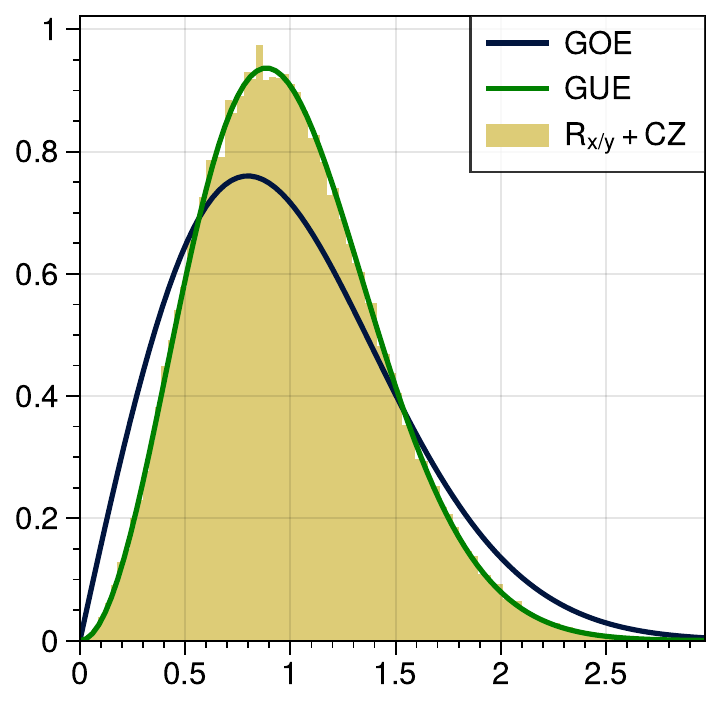}}  
        \hfill
\subfloat[\centering $R_y$ + CP]{
        \centering
        \includegraphics[height=4.0cm]{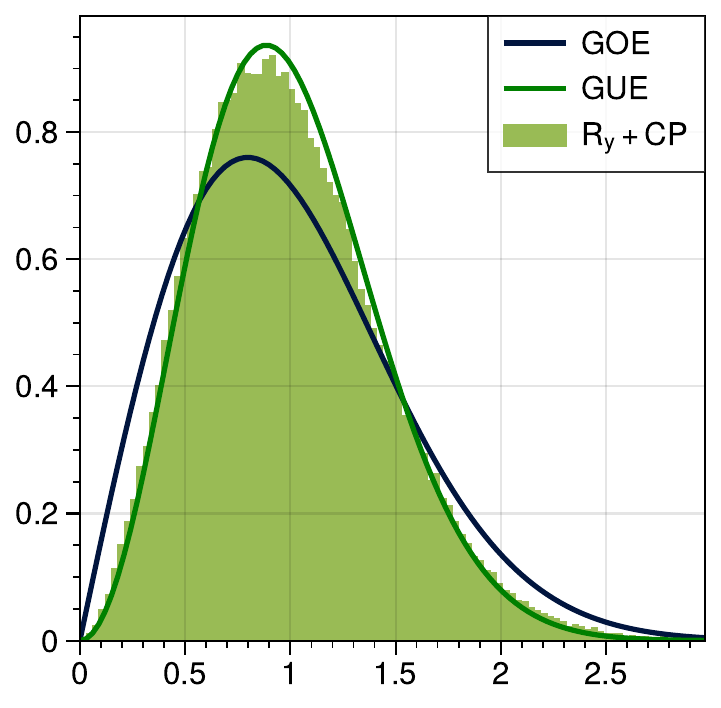}}
\caption{Level spacing distribution at $L=30$ number of layers. We see clearly the agreement with the RMT
distributions. $R_x$ + CZ and $R_y$ + CZ follow GOE, while $R_x$ + CZ+$R_y$ + CZ and $R_y$ + CP
follow GUE.
}
\label{fig:ls30}
\end{figure*}

\begin{figure*}[t]
\centering
\subfloat[\centering $R_x$ + CZ]{
        \centering
        \includegraphics[height=4.0cm]{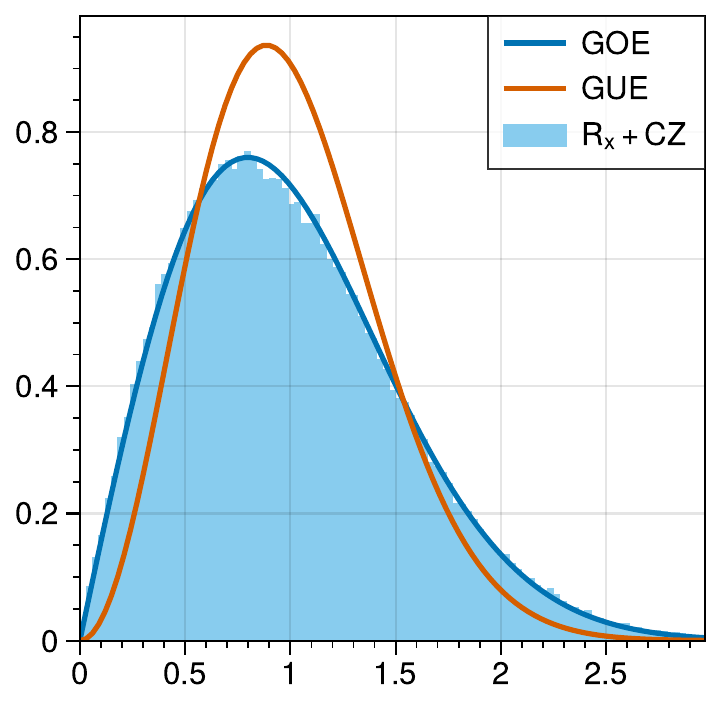}}
        \hfill
\subfloat[\centering $R_y$ + CZ]{
        \centering
        \includegraphics[height=4.0cm]{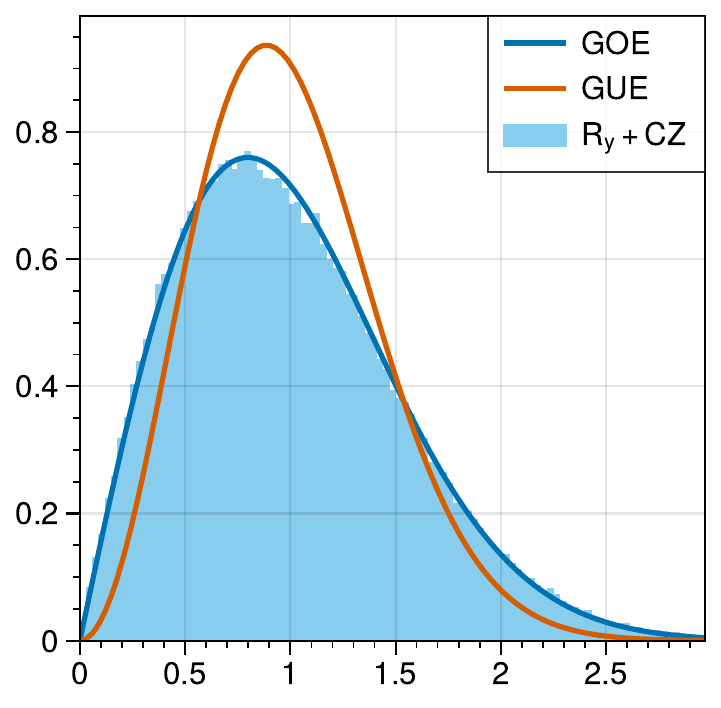}}  
        \hfill
\subfloat[\centering $R_x$ + CZ+$R_y$ + CZ]{
        \centering
        \includegraphics[height=4.0cm]{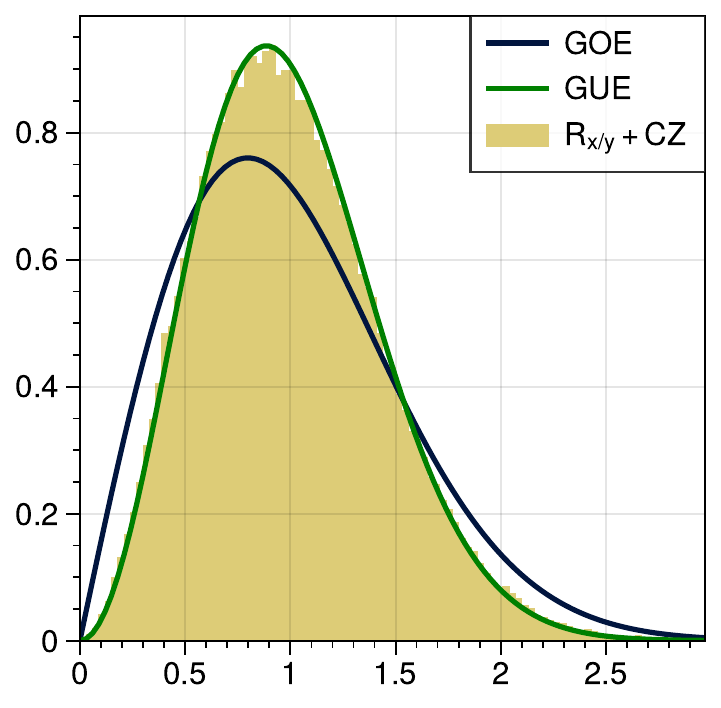}}  
        \hfill
\subfloat[\centering $R_y$ + CP]{
        \centering
        \includegraphics[height=4.0cm]{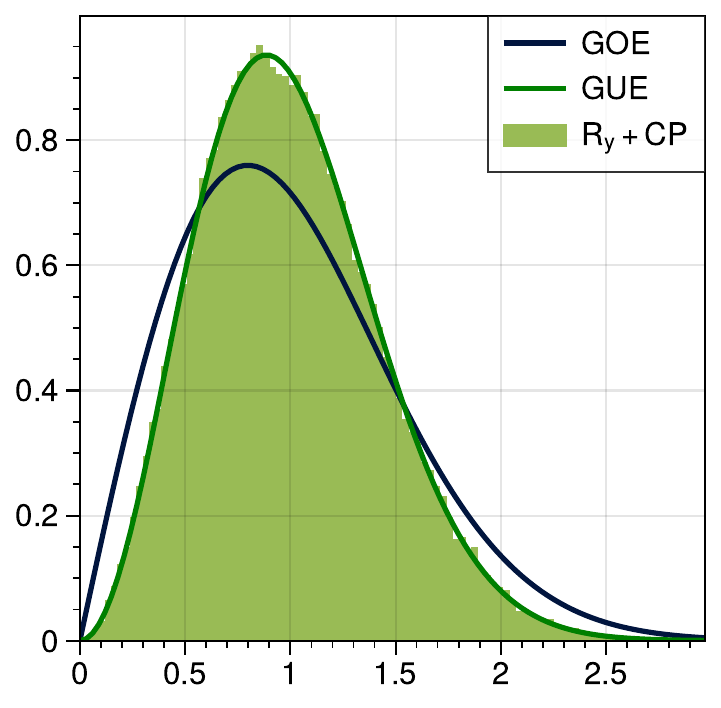}}
\caption{Level spacing distribution at $L=250$ number of layers. We see a precise agreement with the RMT
distributions. $R_x$ + CZ and $R_y$ + CZ follow GOE, while $R_x$ + CZ+$R_y$ + CZ and $R_y$ + CP
follow GUE.}
\label{fig:ls250}
\end{figure*}

For different types of circuit unitaries defined in Section~\ref{sec:vqa} with $L=10$, $30$ and $250$ layers, we collect $500$ random circuit states and draw the corresponding level spacing distributions in Figures~\ref{fig:ls10}--\ref{fig:ls250}.
The modular Hamiltonian of shallow circuit states at $L=10$ displays a clear departure from the RMT predictions, manifesting a lack of level repulsion.
Such distinction is particularly pronounced for the $R_y + \text{CP}$ unitary circuit. However, the emergence of random matrix structure becomes evident as stacking more circuit layers. The agreement between empirical level spacing distributions of random circuit states and RMT predictions \eqref{Wigner} is already quite obvious at $L=30$ and further improved at $L=250$. Note that different choices of unitary gates lead to the emergence of different random matrix ensembles. We observe GUE for the $R_x + \text{CZ}+ R_y + \text{CZ}$ and $R_y + \text{CP}$ circuit unitaries and GOE for $R_x + \text{CZ}$ and $R_y + \text{CZ}$ unitaries. What universality class high-depth random circuit states belong to can be traced from the characteristics of their modular Hamiltonians, or even primarily, from their full density matrices.

We remark that, though the empirical level spacing distribution follows the random matrix theory and exhibits chaotic properties at $L=30$, the entanglement entropy and operator spreading coefficient have not reached saturation and the VQA optimization works smoothly. This points out a difference between the information scrambling measures based on eigenvalues vs. their spacings of modular Hamiltonians.

\subsection{r-statistics}

The previous analysis of the level spacing distribution depends on unfolding the energy spectrum, which is only heuristically defined and has some arbitrariness. Therefore, it would be desirable to have additional diagnostics of quantum chaos that bypass the unfolding procedure. The $r$-statistics, first introduced in \cite{huse_r_stat}, is such a diagnostic tool for short-range correlations, defined without the necessity to unfold the spectrum.

Given the level spacings $s_i$, defined as the differences between adjacent eigenvalues $\cdots < E_i < E_{i+1} < \cdots$ without unfolding, one defines the following ratios:
\begin{align}
\label{eq:ri_definition}
& r_i = \frac{{\rm Min}(s_i, \, s_{i+1})}{{\rm Max}(s_i, \, s_{i+1})} \ , 
\end{align}
which are by definition positive numbers between 0 and 1.
The ratios $r_i$ take very specific values if the energy levels are the eigenvalues of random matrices: For matrices in GOE, GUE and GSE, the ratios are $r_i \approx 0.53590$, $r_i \approx 0.60266$ and $r_i \approx 0.67617$, respectively.
The values become typically smaller for integrable Hamiltonians, approaching $r_i \approx 0.38629$ for a pure Poisson process \cite{r_ratios_theory_values}.

From their very definition, we see that the ratios \eqref{eq:ri_definition} do not require to unfold the spectrum
since their dependence on the local density of states is cancelled by taking the ratio between spacings.
Moreover, each $r_i$ depends on just three adjacent energy levels, rendering it as a sharp microscopic probe of the chaotic/integrable behavior in a small cluster of spectral values.

\begin{figure*}[t]
\centering
\subfloat[\centering The ratios $\{r_i\}_i$ at $L=10$]{
        \centering
        \includegraphics[height=4.5cm]{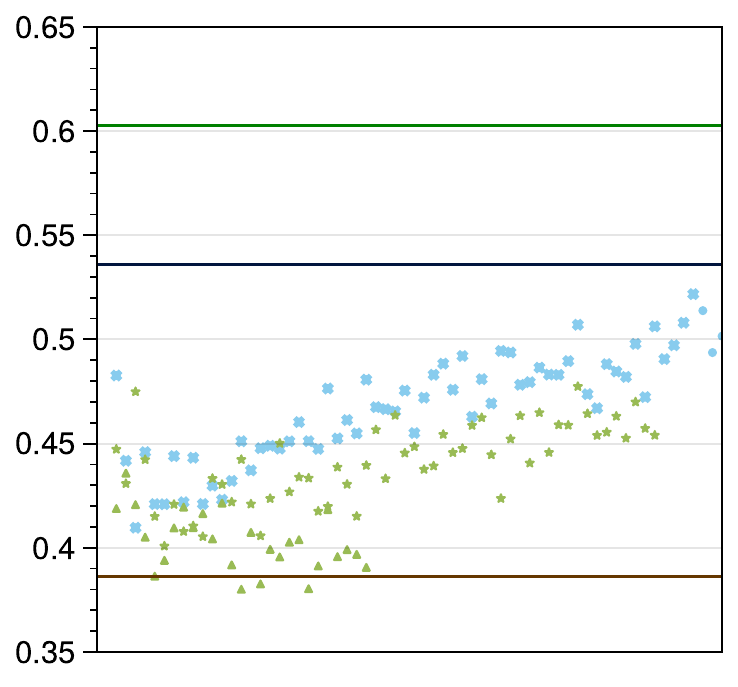}}
\subfloat[\centering The ratios $\{r_i\}_i$ at $L=30$]{
        \centering
        \includegraphics[height=4.5cm]{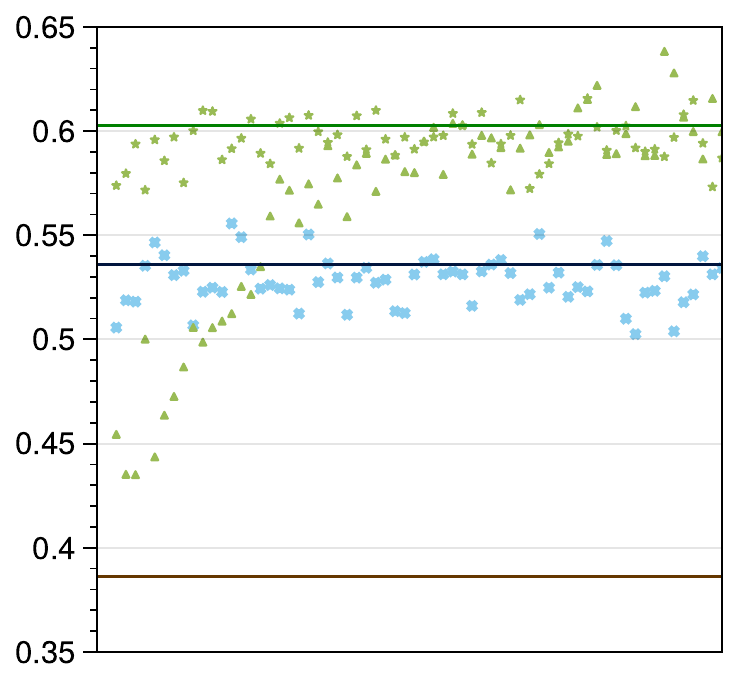}}  
\subfloat[\centering The ratios $\{r_i\}_i$ at $L=250$]{
        \centering
        \includegraphics[height=4.5cm]{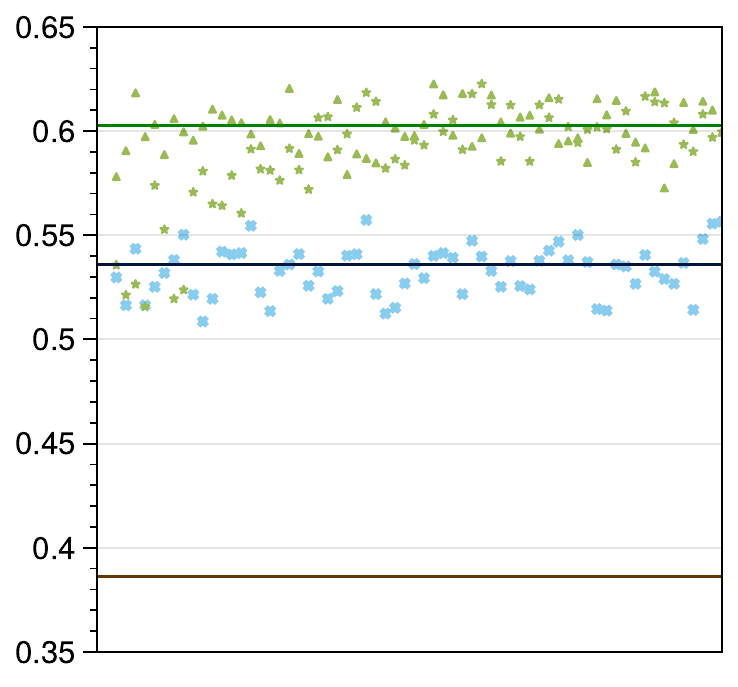}}  
\raisebox{1.cm} {\includegraphics[height=3.0cm]{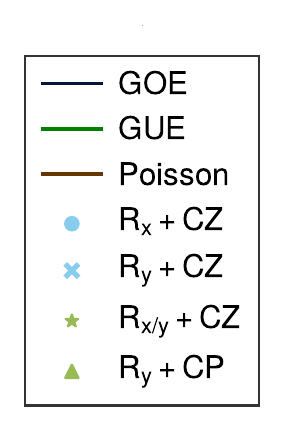}}
\caption{The r-statistics, for a system with 12 qubits, with equal partitions and with $10$ (left), $30$ (center) and $250$ (right) layers.
We observe evident deviations from the expected RMT values for $10$ layers, while a substantial agreement with the RMT predictions is obtained at $30$ layers.
Interesting, the R$_y + $CP circuit stills show deviations from the RMT values in the low lying modes which disappear at $250$ layers.}
\label{fig:r-statistics_eqpart}
\end{figure*}

Here we use the $r$-statistics to study the chaotic properties of the entanglement spectra
as a function of the number of circuit layers $L$.
Under the equal partitioning of $n=12$ qubits and with $L = 10, \, 30, \,250$ layers, the numerical values of $\{r_i\}$ are shown in Figure~\ref{fig:r-statistics_eqpart} where 
we observe the transition from Poisson-like to RMT-like values. 
The low-level eigenstates of the reduced density matrix are more prone to keep their integrable behavior until a sufficient number of entangling layers $L \gtrsim 30$ is reached, where we find the universal GOE/GUE chaotic structure in agreement with the level spacing distribution analysis.

\subsection{Spectral Form Factor} 
\begin{figure*}[t]
\centering
\subfloat[\centering $R_x$ + CZ]{
        \centering
        \includegraphics[height=5.0cm]{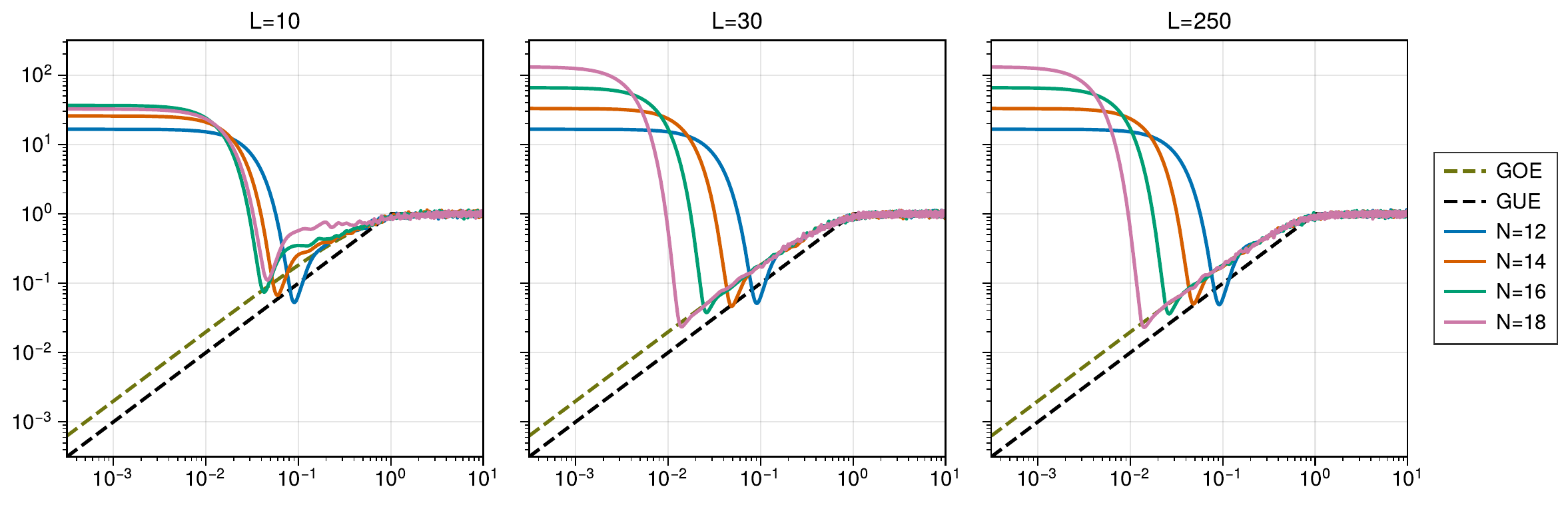}}\hfill
\subfloat[\centering $R_y$ + CZ]{
        \centering
        \includegraphics[height=5.0cm]{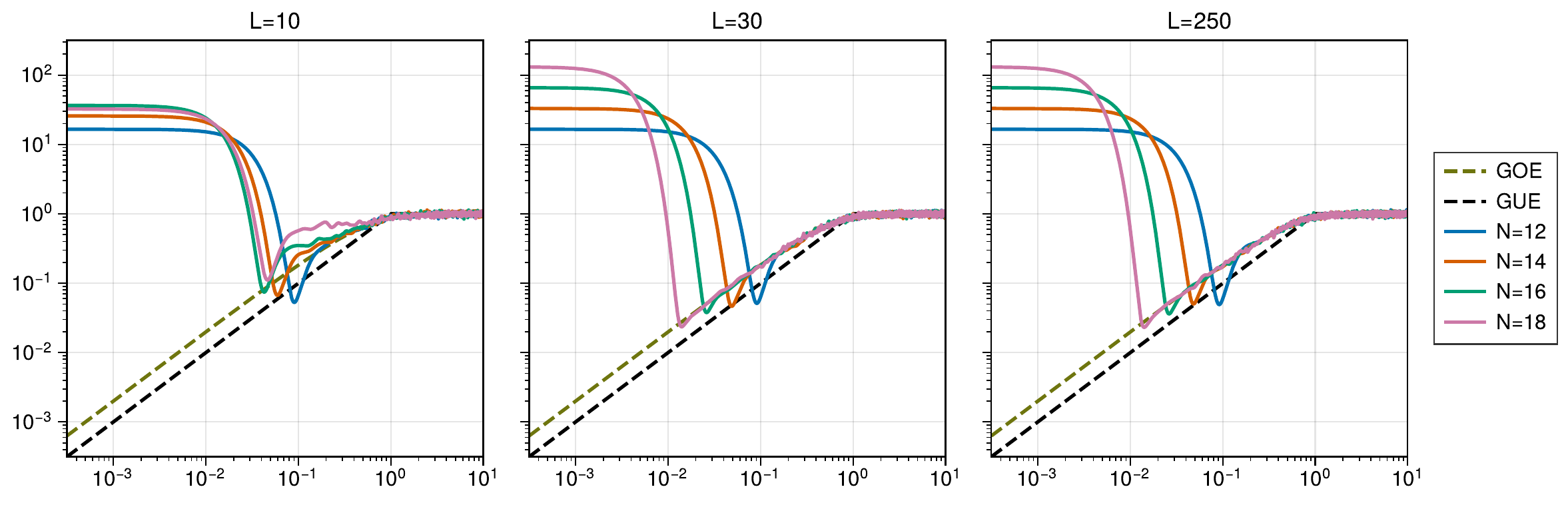}}  
        \hfill
\subfloat[\centering $R_x$ + CZ+$R_y$ + CZ]{
        \centering
        \includegraphics[height=5.0cm]{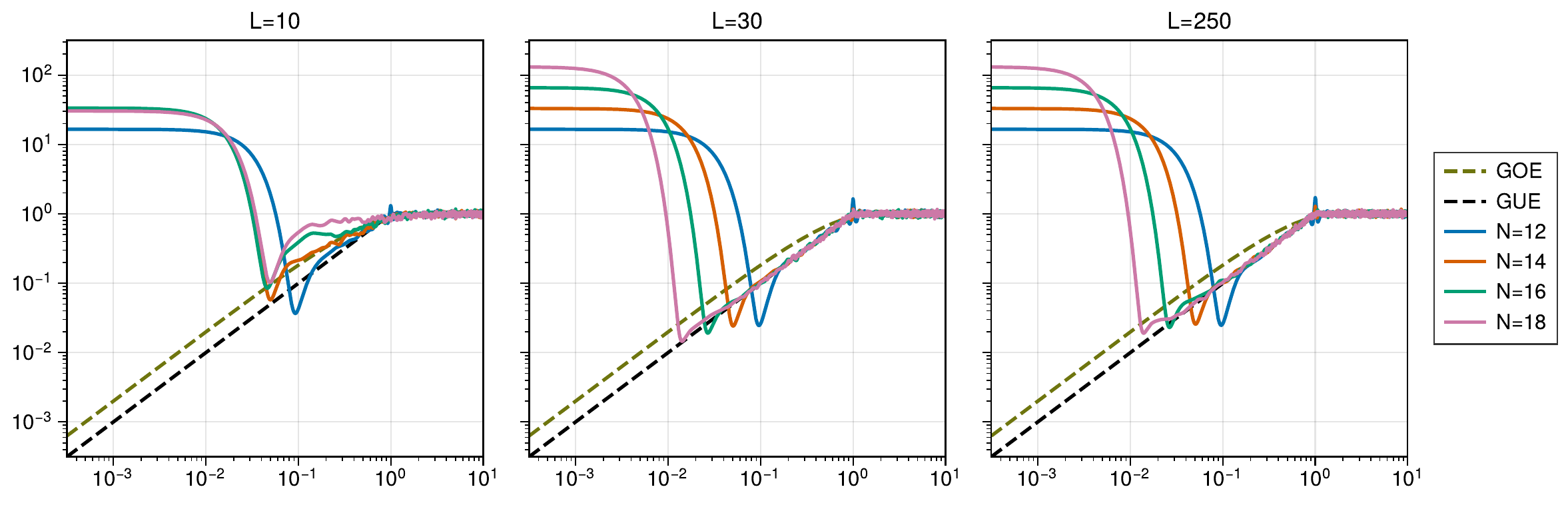}}  
        \hfill
\subfloat[\centering $R_y$ + CP]{
        \centering
        \includegraphics[height=5.0cm]{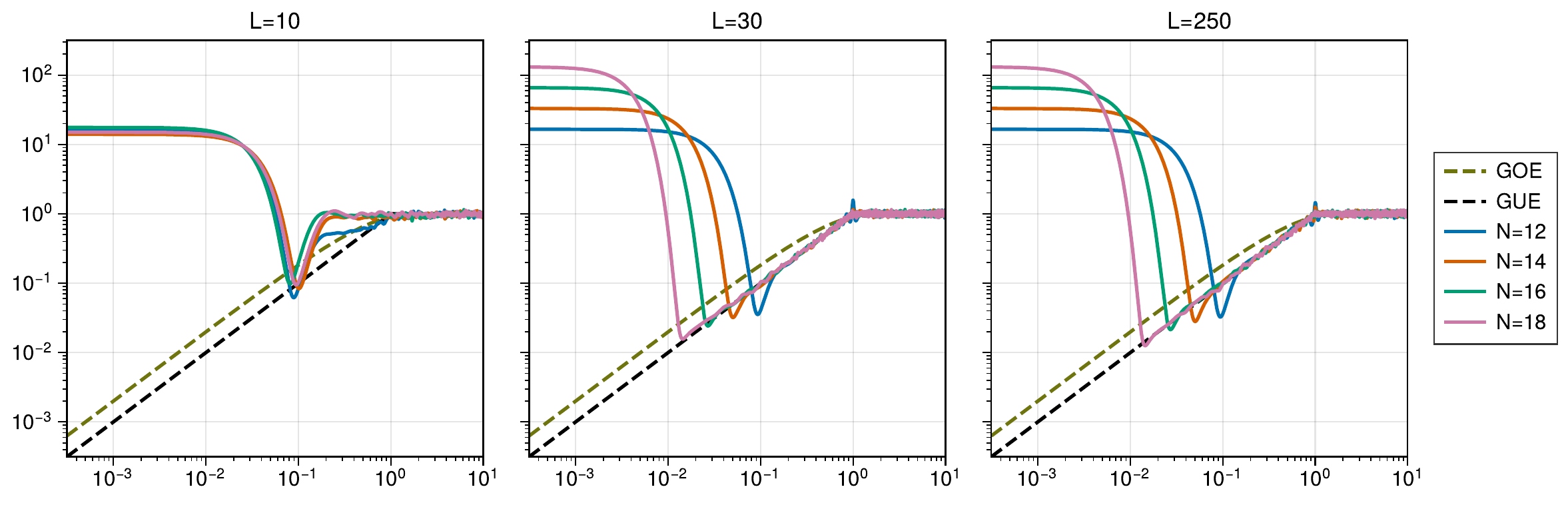}}
\caption{SFF for $10$ (left subfigures), $30$ (central subfigures) and $250$ (right subfigures) layers.
In all cases, the SFFs are very different from the RMT predictions at $10$ layers, while starting from $30$ layers the agreement is excellent.
In particular, the Thouless time, $\tau_{\mathrm{Thoul}}$, clearly decreases while increasing the system size.
As mentioned in the main text, such a behavior is a signal of the ergodic character of the circuits under investigation.}

\label{fig:sff250}
\end{figure*}


The spectral form factor (SFF) is the Fourier transform of the spectral two-point correlation function \cite{mehta_book}.
It can be viewed as a long-range observable, since it probes the agreement of a given unfolded spectrum with RMT at energy scales much larger than the mean level spacing. In particular, SFF can detect the presence of spectral rigidity, and is thus 
a complementary probe of quantum chaos  to the distribution of level spacing and the $r$-statistics that are
short-range observables.
Formally, one defines the analytically continued partition function
\begin{equation}
Z(\tau) = \Tr e^{-i \tau H(\rho_A)} \ ,    
\end{equation}
and the spectral form factor is \cite{Shenker_et_al_SFF_first}
\begin{align}
    K(\tau) = {|Z(\tau)|^2}/{Z(0)^2}.
\end{align}

For a concrete numerical evaluation, we will take the following expression as a robust definition of the spectral form factor \cite{vidmar_1905.06345}: 
\begin{equation}
    \label{eq:SFF_def}
    K(\tau) \equiv \frac{1}{Z} \left\langle \abs{\sum_{i} \rho(e_i) e^{- i 2 \pi e_i \tau}}^2 \right\rangle \ .
\end{equation}
where $e_i$ is the unfolded spectrum of the modular Hamiltonian. 
The normalization factor $Z =  \sum_i \abs{\rho(e_i)}^2$ is chosen to ensure that $K(\tau) \approx 1$ in the limit $\tau \to \infty$.
The bracket $\left\langle \cdots \right\rangle$ denotes the ensemble average over distinct random circuit realizations.
$\rho(e_i)$ is a Gaussian filter  \cite{Shenker_et_al_SFF_second},
\begin{equation}
    \label{eq:gaussian_filter_def}
    \rho(e_i) \equiv \exp{- \frac{ 2 \left(e_i - \bar e\right)^2}{\Gamma^2}} \, ,
\end{equation}
where $\bar e$ and $\Gamma^2$ denote the mean energy and the variance for each unfolded spectrum.
Its purpose is to guarantee that the SFF is mainly affected by eigenvalues located
around the mean value of each unfolded spectrum.

The SFF can be computed analytically for the Gaussian ensembles (GOE and GUE). It reads in the thermodynamic limit as \cite{mehta_book}
\begin{align}
    \label{eq:SFF_gaussian_formula}
    & K_{\mathrm{GOE}}(\tau) = 2\tau - \tau \, \mathrm{ln}(1 + 2\tau) \, , \nonumber \\
    & K_{\mathrm{GUE}}(\tau) = \tau ,
\end{align}
when $0 < \tau < 1$ and $K(\tau) = 1$ when $\tau \geq 1$. Its constancy for $\tau \geq 1$ simply comes from the discreteness of the spectrum and carries no information about spectral correlations.
In particular, since the mean level spacing $\Delta$ is by construction equal to $1$ in the unfolded spectrum, the relevant time scale at which the discreteness of the spectrum becomes relevant is accordingly $\tau \approx 1/\Delta \approx 1$.
This scale is usually called the Heisenberg time, $\tau_{\mathrm{Heis}}$.

The emergence of the random matrix structure in spectral correlations must be investigated for times shorter than the Heisenberg time, $\tau \leq 1$.
The timescale that characterizes the ergodicity of a dynamical system is called the Thouless time, $\tau_{\mathrm{Thoul}}$, defined as the time when the SFF of the dynamical system converges to the universal RMT computation.
More concretely, it is indicated by the onset of the universal linear ramp as in \eqref{eq:SFF_gaussian_formula}. One expects $\tau_{\mathrm{Thoul}}$ to decrease by increasing the system size, in ergodic systems, to approach $0$ in the thermodynamic limit. In contrast, non-ergodic systems show the absence of linear ramp, \textit{i.e.} $\tau_{\mathrm{Thoul}} \sim \tau_{\mathrm{Heis}} \sim 1$, or more generally, unclear scaling of $\tau_{\mathrm{Thoul}}$ with respect to the system size.

Following the above discussion, we computed the empirical SFF for different circuit architectures with $L=10,\, 30$ and $250$ layers, where the ensemble average is replaced with averaging over 50 random circuit samples. See Figure~\ref{fig:sff250}. Inspecting the Thouless time as expanding the system from $n = 12$ to $18$ reveals clear indications of ergodicity breaking at $L=10$, but an expected ergodic behavior for $L=30$ and $250$ layers. They are consistent with the conclusion obtained through the short-range observables in previous subsections.

The circuit reduced density matrix is a random matrix by construction. And thus, it may not be surprising
that the modular Hamiltonian eigenspectrum exhibits chaotic properties of RMTs. It is interesting, however, to trace the reasons for the GOE and GUE structures to the form of the quantum gates.
While random circuit states should generically be in the GUE class, the choice of the gates may generate a modular Hamiltonian whose matrix elements are not  complex-valued, but rather
real or pure imaginary. In such cases, e.g., for $R_x + \text{CZ}$ and $R_y + \text{CZ}$ unitaries, the  eigenspectrum of the corresponding modular Hamiltonian must belong to GOE. 

Note that although the level spacing diagnostics show apparent RMT properties at $L=30$, the local  search of optimal circuit parameters still operates well. It indicates that unlike the diagnostic measures based on  eigenvalues of the modular Hamiltonian, e.g., entanglement entropies and operator spreading coefficients, the quantum chaos diagnostics
constructed from the level spacing of eigenvalues are not  precisely correlated with the efficiency of optimizing control variables.





\section{Discussion and Outlook}
\label{sec:discussion}

We analyzed the universal chaotic properties of random quantum circuits at different depths and how it correlates to the optimization performance of control variables. Our main focus was on the operator spreading and the level spacing distribution for the eigenspectrum of reduced density matrices. We found that the random circuit wavefunction exhibits the chaotic structure of Gaussian matrix ensembles, which can be either GOE or GUE depending on the type and arrangement of unitary gates. 

By changing the direction of the magnetic field coupled to the Ising Hamiltonian used in the VQE experiments, we observed the failure of specific GOE-type variational circuits in reaching the ground state. It suggests the expressibility of variational circuits is not determined alone by their capability of creating highly-entangled states.


Both chaos and entanglement follow from the eigenspectrum structure of the reduced density matrix. 
However, while entanglement and operator spreading are captured by quantities constructed from eigenvalues themselves, there are other measures of quantum chaos beyond the operator spreading which instead relate to their level spacings. We found that the quantum chaos diagnosed by the eigenvalue spacings typically emerges with fewer circuit layers than to come close to the maximum entanglement of random circuit states, which hinders an effective search of optimal circuit variables  \cite{McClean2018bp, cost-dep-bp,entanglement-bp,highdepth,Kim:2021ffs}. 


Such study points to some mismatch between two distinct definitions of quantum chaos, i.e., the BGS conjecture vs.~the operator spreading measured by OTOC. Note that the random circuit exhibits the BGS-type chaotic structure before reaching the complete spreading of operators in OTOC. To the best of our knowledge, this is the first example of a genuine many-body setup in which such discrepancy is observed. Previous studies dealt only with single-body examples with classical counterparts
 \cite{bunimovich, scaffidi_2020, benenti_2020, lucas_2021, benenti_2021, kirby_2021, kidd_2021}.


As for future studies, it would be interesting to explore the connection between the graph structure of variational circuits, their effectiveness as the eigensolver of distinct Hamiltonians, and the emergence of quantum chaos in random circuit states.
A popular measure of information mixing is the $k$-design state that cannot be distinguished
from the Haar random state when considering averages
of polynomials of degree not higher than $k$. It would be useful to investigate the
relationship in the framework of random quantum circuits 
between the $k$-design structure and the quantum chaos measures that we analyzed. Some results in this direction have been investigated in \cite{haferkamp2020quantum}.

Another intriguing line of investigation is to study the non-stabilizerness of variational circuits --- often referred to in the literature as \textit{magic} and regarded to be the source of quantum advantage in many computing problems \cite{Gottesman:1998hu, knill_theorem}. An explicit measure of magic was recently proposed in \cite{hamma_magic_measure}. Its relations with quantum chaos was studied in \cite{hamma_quantum_quantum}. It would be interesting to better investigate the role of magic in the VQA problems, following \cite{hamma_tgates_2020}.

\section*{Acknowledgments}
We would like to thank Thi Ha Kyaw, Alexey Milekhin and Jan Olle for valuable discussions. The work of J.K. is supported by the NSF grant PHY-1911298 and the Sivian fund.
The work of Y.O. is supported in part by the Israeli Science Foundation Center
of Excellence.
DR acknowledges the support by the Institute
for Basic Science in Korea (IBS-R024-Y2
 and IBS-R024-D1).

 \providecommand{\href}[2]{#2}\begingroup\raggedright\endgroup

\end{document}